\documentclass[review,12pt]{elsarticle}

\usepackage{amssymb}
\usepackage{amsmath}
\usepackage{threeparttable}
\usepackage{subcaption}
\usepackage{booktabs}
\usepackage[ruled,vlined]{algorithm2e} 
\usepackage{xcolor}
\usepackage{float}

\usepackage[colorlinks=true, linkcolor=red, anchorcolor=blue,citecolor=green]{hyperref}

\journal{Journal of Computational Physics}

\begin{document}

\begin{frontmatter}



\title{Applying contact angles based on a continuously moving contact line in 2D VOF simulations.}


\author[label1,label2]{Tian-Yang Han} %% Author name
\ead{hantianyang@ucas.ac.cn}
\author[label2]{Jie-Yun Pan\corref{cor1}} %% Author name
\ead{yi.pan@sorbonne-universite.fr}
\author[label3]{Chong-Sen Huang} %% Author name
\ead{cshuang@stu.xjtu.edu.cn}
\author[label3]{Jie Zhang} %% Author name
\ead{j_zhang@xjtu.edu.cn}
\author[label1]{Ming-Jiu Ni\corref{cor2}} %% Author name
\ead{mjni@ucas.ac.cn}
\author[label2]{Stéphane Popinet} %% Author name 
\ead{popinet@basilisk.fr}
\author[label2,label4]{Stéphane Zaleski} %% Author name
\ead{stephane.zaleski@sorbonne-universite.fr}
\cortext[cor1]{Corresponding author}
\cortext[cor2]{Corresponding author}

\affiliation[label1]{organization={School of Engineering Science, University of Chinese Academy of Sciences},
	% city={},
	postcode={101408},
	state={Beiijng},
	country={China}}

\affiliation[label2]{organization={Sorbonne Université, CNRS, Institut Jean Le Rond d'Alembert UMR 7190},
	% city={},
	postcode={F-75005},
	state={Paris},
	country={France}}

\affiliation[label3]{organization={State Key Laboratory for Strength and Vibration of Mechanical Structures, School of Aerospace, Xi’an Jiaotong University},
            city={Xi’an},
            postcode={710049},
            state={Shaanxi},          
            country={China}}

\affiliation[label4]{organization={Institut Universitaire de France},
	% city={},
	state={Paris},          
	country={France}}  

%% Abstract
\begin{abstract}
A height-function-based numerical approach is developed for enforcing contact angles on flat and curved solid surfaces within two-dimensional volume-of-fluid simulations. This method incorporates the contact line position into the curvature estimation in the contact line cell, where the interface normal is constrained to the prescribed angle to ensure smooth contact line motion. On flat solid surfaces, the proposed model achieves higher accuracy than the conventional vertical height-function method for enforcing very small and very large contact angles. Moreover, it extends naturally to curved solid surfaces represented by the embedded boundary method, enabling the imposition of arbitrary contact angles without inducing significant spurious currents near the contact line. Numerical validations confirm the accuracy and robustness of the proposed method, demonstrating its applicability to multiphase flows involving complex wall geometries and contact-line dynamics.
\end{abstract}


% \begin{highlights}
% \end{highlights}

%% Keywords
% \begin{keyword}



% \end{keyword}

\end{frontmatter}



\section{Introduction}

Multiphase flows involving moving contact lines are ubiquitous in nature and industrial applications, such as ink-jet printing \cite{soltman2013inkjet}, enhanced oil recovery \cite{alvarado2010enhanced}, and curtain coating \cite{fullana2020dynamic}. Numerical simulation of such flows faces three major challenges: (1) accurate capturing of the multiphase interface, (2) consistent discretization of the surface-tension force, and (3) reliable modeling of the fluid–solid interaction. The last aspect primarily concerns  contact line dynamics, which inherently give rise to pressure and shear-stress singularities when the no-slip boundary condition is imposed on the solid wall \cite{snoeijer2013moving}. To overcome this difficulty, an appropriate contact-line model must be introduced, typically combining a moving mechanism for the contact line, such as Navier slip, with a prescribed contact-angle condition to reflect wall wettability \cite{sui2014numerical}. Over the past decades, a variety of interface-capturing methods—including Front Tracking (FT), Level Set (LS), Volume-of-Fluid (VOF), and diffuse-interface (DI) approaches—have been employed to simulate moving contact line problems \cite{manservisi2009variational,spelt2005level,renardy2001numerical,afkhami2009mesh,dupont2010numerical,ding2007wetting}. Among these, the VOF method is particularly attractive due to its excellent mass conservation and its natural ability to handle interface coalescence and breakup \cite{scardovelli1999direct}, and is therefore adopted in the present study.

Within the VOF framework, consistent discretization of surface-tension forces can be achieved by combining height-function (HF) based curvature estimation with the balanced-force continuum surface force (CSF) formulation \cite{popinet2018numerical}. Contact-line motion is enabled by the inherent implicit slip of the VOF method \cite{afkhami2009mesh}. To impose a prescribed contact angle, two main strategies have been proposed. The first assigns the volume fraction in ghost cells outside the computational domain \cite{renardy2001numerical,dupont2010numerical}, while the second relies on the HF approach, in which height functions are specified either in ghost cells or in neighboring non-interfacial cell columns \cite{afkhami2008height,afkhami2009height}. Both approaches enforce the desired contact angle by adjusting the local gradient of the volume fraction or height function at the contact line. In this work, we adopt the HF-based approach owing to its superior accuracy in curvature estimation \cite{cummins2005estimating}. In 2D, the HF-based contact-angle model of Afkhami and Bussmann \cite{afkhami2008height} consists of two parts: a horizontal HF scheme applicable for contact angles between $45^{\circ}$ and $135^{\circ}$, and a vertical HF scheme for angles outside this range. While the horizontal HF model can achieve first-order convergence in curvature estimation at the contact line \cite{han2021consistent}, the vertical HF model fails to converge.

\begin{table}[t]
\centering
\begin{threeparttable}
\caption{Comparison of contact-angle models for complex geometries within the geometric VOF framework combined with EBM solid representation.}
\label{tab:comparison}
\begin{tabular}{l p{9.5cm}}
\hline
          & \textbf{Contact-angle enforcement approach} \\
\hline
Tavares et al.~\cite{tavares2024coupled} & Linear extrapolation of the reconstructed interface within mixed cells to assign ghost-cell volume fractions in the solid region. \\
Chen et al.~\cite{chen2025volume}\tnote{a} & Refined ghost-cell extrapolation of Tavares’ method by first identifying the contact-line cell. \\
Huang et al.~\cite{huang20252d}\tnote{a} & Parabola fitting of height functions near the contact line to reconstruct interface geometry and extrapolate it into the solid region. \\
\textbf{Present study}\tnote{a}          & Curvature estimation based on the identified contact-line position within the contact-line cell, without any extrapolation into the solid region. \\
\hline
\end{tabular}
\begin{tablenotes}
\footnotesize
\item[a] These methods account for the presence of the solid phase in mixed cells during interface reconstruction and geometric advection.
\end{tablenotes}
\end{threeparttable}
\end{table}

The above-mentioned contact-angle models were developed for flat solid boundaries. For contact-line motion on solid surfaces with complex geometries, however, accurately capturing the fluid–solid interaction becomes more challenging. Two main strategies have been proposed: (i) body-fitted grids \cite{fan2013piecewise} and (ii) immersed or embedded boundary methods (IBM/EBM) implemented on Cartesian grids \cite{patel2017coupled,o2018volume,tavares2024coupled,huang20252d}. In the present work, we focus on the latter. Patel et al.~\cite{patel2017coupled} and O’Brien et al.~\cite{o2018volume} coupled IBM with algebraic or geometric VOF schemes, wherein the interface normal $\mathbf{n}$ at the contact line is prescribed to satisfy the desired contact angle and the curvature is then computed as $\kappa=\nabla\cdot\mathbf{n}$. While effective in imposing the required contact angle, this implementation may suffer from local mass-conservation errors and limited grid convergence in curvature estimation \cite{cummins2005estimating}.

More recent developments have combined EBM with geometric VOF, enabling sharp representations of both fluid–fluid and solid–fluid interfaces while evaluating curvature via the more accurate HF method. Tavares et al.~\cite{tavares2024coupled} imposed the contact angle by extending the reconstructed interface in mixed cells into the solid region to assign ghost-cell volume fractions; however, this strategy deteriorates for very small or very large contact angles. Chen et al.~\cite{chen2025volume} improved this strategy by extending the reconstructed interface only within the contact-line cell, leading to more accurate ghost-cell volume fractions and corresponding height functions. In addition, Chen et al. also take into account the presence of the solid phase within mixed cells during interface advection—an effect neglected in the method of Tavares et al. Nevertheless, as pointed out by Huang et al.~\cite{huang20252d}, the divergence-free constraint in mixed cells is not strictly satisfied in Chen et al.’s formulation, which may still introduce local mass-conservation errors. To address this issue, Huang et al. proposed a conservative advection scheme for mixed cells and developed a more sophisticated parabola-fitting procedure to directly assign height functions near the contact line, thereby achieving robust accuracy over the full range of contact angles. The differences among these methods, implemented within the geometric VOF–EBM framework, are summarized in Table~\ref{tab:comparison}, together with the present approach for comparison.

Despite these advances, existing contact-angle models within VOF simulations still face some limitations. On flat solid surfaces, the conventional vertical HF model—intended for very small and very large contact angles—fails to achieve the same accuracy as the horizontal HF formulation. For solid surfaces with complex geometries, although the parabola-fitting method of Huang et al. \cite{huang20252d} demonstrates good accuracy, its implementation remains relatively complex. In this work, we propose a HF-based contact-angle model that incorporates the contact-line position into the curvature estimation within contact-line cells. The objectives of the proposed model are threefold: (i) to improve the accuracy of enforcing very small and very large contact angles on flat solid surfaces compared with the conventional vertical HF model; (ii) to enable a straightforward extension of the model from flat to curved solid boundaries while remaining simple to implement; and (iii) to ensure accurate curvature estimation near the contact line across the full range of contact angles, while minimizing spurious currents.

\section{Numerical method}
\label{sec2}

The numerical methods employed in this paper are developed based on the open-source code $Basilisk$ \cite{popinet2015quadtree}, which solves the incompressible Navier-Stokes equations on an adaptive Cartesian mesh using a time-staggered approximate projection method. In what follows, we first provide a brief overview of the numerical schemes implemented in $Basilisk$ for interface tracking, surface tension discretization, and contact angle enforcement. The interface-tracking algorithm is then modified in cut cells to account for the presence of solid boundaries. Finally, we present a new contact-angle model, initially developed for flat solid boundaries to improve accuracy at very small and very large angles, and subsequently extended to complex geometries to impose arbitrary contact angles. The algorithms developed in this work are freely available through the Basilisk sandbox \cite{sandbox}.

\subsection{Governing equations}

Considering the interfacial flow of two immiscible fluids, the dynamics are governed by the incompressible Navier-Stokes equations:
\begin{equation}
  \rho(\frac{\partial \mathbf{u}}{\partial t} + \mathbf{u}\cdot\nabla\mathbf{u}) = -\nabla p + \nabla\cdot(2\mu\mathbf{D}) + \sigma\kappa\delta_s\mathbf{n}
  \label{NSeq}
\end{equation}
\begin{equation}
  \nabla\cdot\mathbf{u} = 0
  \label{nondiv}
\end{equation}
where $\mathbf{u}$, $p$, $\rho$, $\mu$ and $\sigma$ are the fluid velocity vector, pressure, density, dynamic viscosity, and surface tension coefficient, respectively. The strain–rate tensor is defined as $\mathbf{D} = (\partial_j u_i + \partial_i u_j)/2$. The unit normal vector $\mathbf{n}$ and curvature $\kappa$ characterize the interface geometry, while $\delta_s$ is the Dirac distribution function that localizes the surface tension force at the interface. The numerical discretization of the advection and viscous terms in Eq.~(\ref{NSeq}), as well as the pressure–velocity decoupling strategy, follows the method described in \cite{popinet2009accurate} and is not repeated here.

\begin{figure}[tbp]
    \centering
    \begin{subfigure}[b]{0.3\textwidth}
        \centering
        \includegraphics[width=\textwidth]{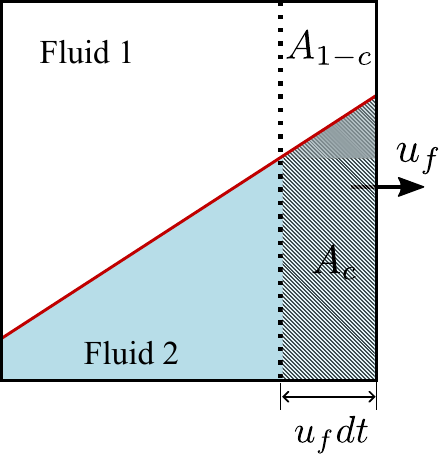}
        \caption{}
        \label{fig:sub0-1-1}
    \end{subfigure}
    \hspace{2cm}
    \begin{subfigure}[b]{0.3\textwidth}
        \centering
        \includegraphics[width=\textwidth]{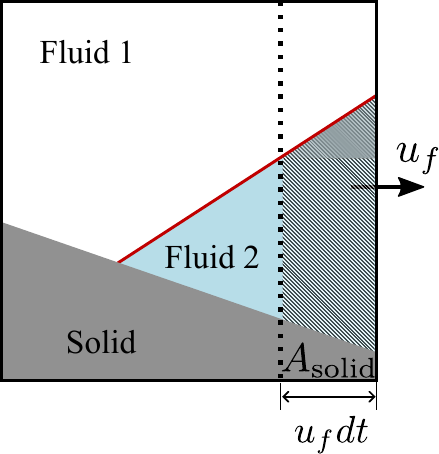}
        \caption{}
        \label{fig:sub0-1-2}
    \end{subfigure}
    \caption{Schematic of linear interface reconstruction and geometric advection flux estimation. (a) Without a solid boundary. (b) With an embedded solid boundary. Here, $u_f$ denotes the fluid velocity at the cell face, and $dt$ the time step. The shaded region indicates the volume of Fluid~2 advected over the distance $u_f dt$. $A_c$, $A_{1-c}$, and $A_{\text{solid}}$ represent the volumes of Fluid~2, Fluid~1, and the solid within the swept region, respectively.}
    \label{fig0-1}
\end{figure}

The interface between the two fluids is captured using the VOF method. A scalar $c$, representing the volume fraction of the reference fluid in each cell, is introduced and its evolution is governed by
\begin{equation}
  \frac{\partial c}{\partial t} + \nabla\cdot(\mathbf{u}c) = 0.
  \label{advc}
\end{equation}
The main challenge in solving Eq.~(\ref{advc}) lies in accurately estimating the advection flux of $c$ across cell faces. In this work, we employ a geometric approach. Specifically, a linear interface segment is reconstructed in each interfacial cell ($0 < c < 1$), from which the advection flux is then computed through simple geometric calculations, as illustrated in Fig.~\ref{fig0-1}(a).

\subsection{Interface tracking in cut cells}
\label{sec:iftrack}

In $Basilisk$, complex solid boundaries are represented using the embedded boundary method (EBM) \cite{johansen1998cartesian}. As shown in Fig.~\ref{fig0-1}(b), the solid phase is embedded within the Cartesian mesh and described by a second volume fraction field $c_s$, with the fluid taken as the reference phase. The corresponding fluid face fractions are denoted by $s_f$. Since the solid boundaries considered here are stationary, $c_s$ and $s_f$ are initialized at the beginning of the simulation, from which the solid boundary is reconstructed as a linear segment within each cut cell containing both solid and fluid.

In this section, we address the reconstruction and advection of the fluid–fluid interface in cut cells. As illustrated in Fig.~\ref{fig0-1}(b), when the solid boundary and the fluid–fluid interface coexist within the same grid cell, both the linear interface reconstruction and the geometric  flux estimation for advection must be modified to account for the presence of the solid phase. Neglecting the solid volume fraction in these procedures can lead to unphysical results. For example, Tavares et al. \cite{tavares2024coupled} ignored the solid phase during the geometric interface advection, which caused an artificial absorption of fluid mass by the solid and consequently violated local mass conservation.

\subsubsection{Linear interface reconstruction}

In the VOF method, linear interface reconstruction represents the fluid–fluid interface within a cell by the equation
\begin{equation}\label{ifeq}
\mathbf{n} \cdot \mathbf{x} = \alpha,
\end{equation}
where $\mathbf{n}$ is the unit normal vector of the interface and $\alpha$ is a constant. The reconstruction requires establishing the following two relationships:
\begin{equation}\label{positive}
c = c(\mathbf{n},\;\alpha),
\end{equation}
\begin{equation}\label{negtive}
\alpha = \alpha (\mathbf{n},\;c).
\end{equation}
For a rectangular cell, these relationships can be derived analytically \cite{scardovelli2000analytical}. In cut cells, however, the situation becomes more complicated due to the irregular shape of the fluid region. To address this, Huang et al. \cite{huang20252d} adopted an interface reconstruction strategy originally developed for arbitrary unstructured grids \cite{dai2018analytical}, while Chen et al. \cite{chen2025volume} proposed a polygon segmentation algorithm to identify the vertices of fluid-occupied polygon, thereby facilitating area computation. Here, we propose an alternative approach that directly exploits the polygonal geometry of the fluid region. In what follows, we first address the \textit{direct problem} \eqref{positive}, namely, computing the volume fraction $c$ from a given interface equation.

\begin{figure}[tbp]
    \centering
    \begin{subfigure}[b]{0.3\textwidth}
        \centering
        \includegraphics[width=\textwidth]{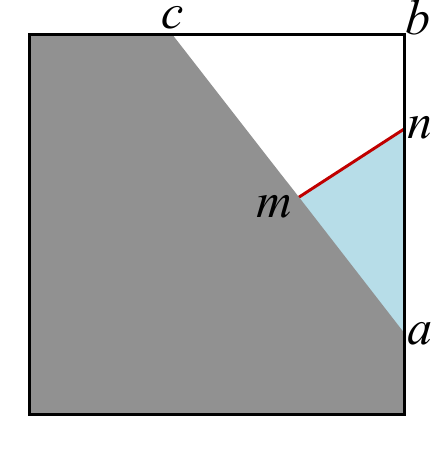}
        \caption{}
    \end{subfigure}
    \hfill
    \begin{subfigure}[b]{0.3\textwidth}
        \centering
        \includegraphics[width=\textwidth]{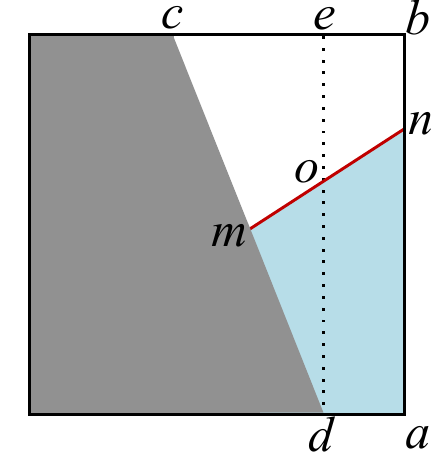}
        \caption{}
    \end{subfigure}
    \hfill
    \begin{subfigure}[b]{0.3\textwidth}
        \centering
        \includegraphics[width=\textwidth]{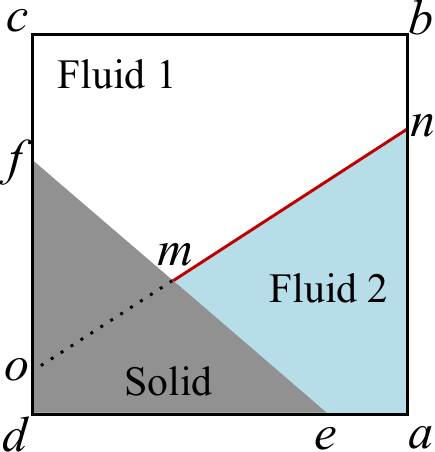}
        \caption{}
    \end{subfigure}
    \caption{Schematic of cut cells with different fluid-region geometries: (a) triangular, (b) quadrilateral, and (c) pentagonal.}
    \label{fig0-4}
\end{figure}

As illustrated in Fig.~\ref{fig0-4}, the fluid region in a cut cell may exhibit three different geometrical configurations. In Fig.~\ref{fig0-4}(a), the fluid region $abc$ forms a triangle. In this case, the volume fraction $c$ of the reference fluid (Fluid 2), corresponding to the area $anm$, can be readily calculated once the intersection points $m$ and $n$ between the interface and the triangular region $abc$ are determined. When the fluid region 
is a quadrilateral, as shown in Fig.~\ref{fig0-4}(b), the area of the reference fluid $anmd$ can be evaluated by partitioning the quadrilateral $abcd$ into a rectangle $abed$ and a triangle $cde$. For these two subregions, analytical expressions are available to compute the corresponding area of the reference fluid, denoted by $omd$ and $odan$, respectively. The total volume fraction $c$ is then obtained by summing the contributions from both subregions. Finally, the fluid region may form a pentagon, as shown in Fig.~\ref{fig0-4}(c). This case is handled by extending the fluid–fluid interface $nm$ to $no$. 
The volume fraction of the reference fluid is then calculated as the difference between (i) the contribution from the entire rectangle $abcd$ defined by the extended interface $no$, corresponding to the area $anod$, and (ii) the contribution from the triangular region $def$ defined by the interface segment $mo$, corresponding to the area $emod$.

\begin{algorithm}[t]
\caption{Bisection iteration for solving \textit{inverse problem} \eqref{negtive}}
\KwIn{Volume fraction $c$, interface normal vector $\mathbf{n}$, vertex coordinates $p_i$ of the polygonal fluid region, tolerance $\varepsilon$}
\KwOut{Interface parameter $\alpha$}

Select appropriate $c=c(\mathbf{n},\alpha)$ according to the polygon shape;

Substitute $p_i$ into the interface equation~\eqref{ifeq} to get corresponding $\alpha_i$;

Evaluate $c_i=c(\mathbf{n},\alpha_i)$ to identify the lower and upper bounds $c_{\min}$ and $c_{\max}$,  along with the corresponding $\alpha_{\min}$ and $\alpha_{\max}$\;

\While{not converged}{
  Compute the midpoint $\alpha_m = (\alpha_{\min}+\alpha_{\max})/2$;
  
  Evaluate $c_m=c(\mathbf{n}, \alpha_m)$\;
  \eIf{$|c_m-c| < \varepsilon$}{
    Convergence reached, \Return $\alpha_m$\;
  }{
    Update $\alpha_{\min}$ or $\alpha_{\max}$ based on the sign of $(c_m-c)$\;
  }
}
\end{algorithm}

After addressing the \textit{direct problem} \eqref{positive}, the remaining task is to solve the \textit{inverse problem} \eqref{negtive}, namely, determining $\alpha$ from the given volume fraction $c$ and interface normal vector $\mathbf{n}$. To this end, a bisection iteration method is employed. The complete procedure is summarized in Algorithm~1.

\subsubsection{Geometric interface advection}

In the discretization of the advection equation Eq.~\eqref{advc}, the advection flux $F_c$ of the volume fraction $c$ across a cell face is given by
\begin{equation} \label{flux}
F_c = \frac{A_c}{\Delta^2},
\end{equation}
where $\Delta$ is the grid size and $A_c$ denotes the area of the reference fluid advected over the distance $u_fdt$, as illustrated by the shaded region in Fig.~\ref{fig0-1}(a). This formulation can be readily extended to cut cells, as shown in Fig.~\ref{fig0-1}(b). In this case, the presence of the solid phase within the swept region (the rectangle to the right of the dotted line) must be taken into account, and Eq.~\eqref{positive} can be employed to compute the area of the shaded region.

However, applying the formulation \eqref{flux} to cut cells may introduce local volume conservation errors. To illustrate this, we first revisit the simpler case of rectangular cells. As shown in Fig.\ref{fig0-1}(a), consider the advection of Fluid~1 (with volume fraction $1-c$) across the right cell face. In this case, the corresponding flux $F_{1-c}$ can be expressed as
\begin{equation} \label{flux2}
F_{1-c} = \frac{A_{1-c}}{\Delta^2},
\end{equation}
where $A_{1-c}$ denotes the area of the Fluid 1 within the swept region. Adding Eqs.~\eqref{flux} and \eqref{flux2} yields
\begin{equation} \label{fluxt}
F_{c} + F_{1-c} = \frac{u_f dt \Delta}{\Delta^2},
\end{equation}
which indicates that the total advected flux of both fluids equals the area of the swept region. Extending this relation to the other three faces of the cell and summing them up gives
\begin{equation} \label{fluxts}
\sum_f (F_{c} + F_{1-c}) = \sum_f \frac{u_f dt \Delta}{\Delta^2} = 0,
\end{equation}
demonstrating that the fluid volume is conserved for the current cell. Here, the divergence-free condition of face velocity $\sum_f u_f = 0$ is employed.

If the above analysis is applied to a cut cell, as illustrated in Fig.~\ref{fig0-1}(b), we obtain
\begin{equation} \label{fluxtsc}
\sum_f (F_{c} + F_{1-c}) = \sum_f \frac{u_f dt \Delta}{\Delta^2} - \sum_f \frac{A_{\text{solid}}}{\Delta^2} \neq 0,
\end{equation}
where $A_{\text{solid}}$ represents the area of the solid phase within the swept region associated with each cell face. This result demonstrates that local volume conservation is violated in the cut cell. It is worth noting that the advection scheme of Chen et al. \cite{chen2025volume} is essentially equivalent to the formulation in \eqref{flux}, and therefore inherits the same volume-conservation issue.

\begin{figure}[t]%% placement specifier
    \centering%% For centre alignment of image.
    \includegraphics[width=0.35\textwidth]{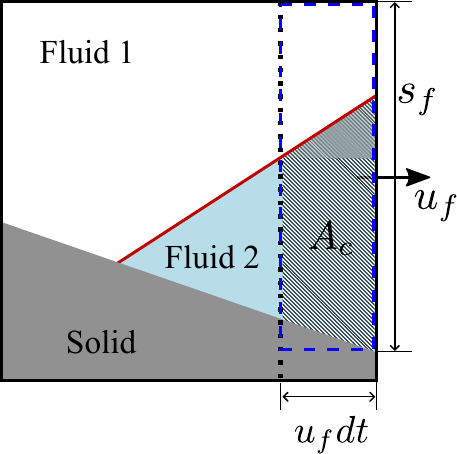}
    \caption{Schematic of the corrected advection flux estimation in a cut cell. Here, $u_f$ denotes the fluid velocity at the cell face, $s_f$ the fluid face fraction and $dt$ the time step. The shaded region $A_c$ represents the advected volume of the reference fluid over the distance $u_f dt$, evaluated geometrically from Eq.~\eqref{positive}.}\label{fig0-5}
\end{figure}

Based on the above analysis, a correction to the flux formulation \eqref{flux} is required to ensure the volume conservation in cut cells. Huang et al. \cite{huang20252d} addressed this issue by introducing a volume-compensation scheme, in which the width of the swept region was adjusted to offset the fluid volume occupied by the solid. In contrast, we propose a simpler remedy that requires only a minor modification to the formulation \eqref{flux}, namely,
\begin{equation} \label{fluxc}
F_c = \frac{A_c}{\Delta^2}\cdot\frac{s_f u_f dt \Delta}{A_c + A_{1-c}},
\end{equation}
where $s_f$ denotes the face fraction occupied by the fluid phase. Compared to Eq.~\eqref{flux}, a correction factor is introduced on the right-hand side of Eq.~\eqref{fluxc}. As illustrated in Fig.~\ref{fig0-5}, the numerator $s_f u_f dt\Delta$ represents the fluid volume to be advected over the distance $u_f dt$ (indicated by the blue dashed box), while the denominator $(A_c + A_{1-c})$ denotes the geometrical fluid volume within the swept region, which can be evaluated from Eq.~\eqref{positive}. Using this new formulation,  and applying the same correction factor to evaluate $F_{1-c}$, Eq.~\eqref{fluxts} can then be rewritten as
\begin{equation} \label{fluxts2}
\sum_f (F_{c} + F_{1-c}) = \sum_f \frac{s_f u_f dt \Delta}{\Delta^2} = 0,
\end{equation}
which demonstrates that the fluid volume is conserved in the cut cell. Here, the divergence-free condition for the face velocity of the cut cell is employed, namely $\sum_f s_f u_f = 0$.

\subsection{Surface tension discretization}

The surface tension force in Eq.~(\ref{NSeq}) is evaluated using the continuum surface force (CSF) model \cite{brackbill1992continuum},
\begin{equation}
  \sigma\kappa\delta_s\mathbf{n} = \sigma\kappa\nabla c.
\end{equation}
In the numerical implementation, it is essential that the discretization of the gradient operator be consistent with that used for the pressure $p$ in Eq.~(\ref{NSeq}), in order to minimize spurious currents \cite{popinet1999front}. This approach is commonly referred to as the balanced-force CSF scheme \cite{francois2006balanced}.

The remaining issue in the discretization of the surface tension force is the estimation of the curvature $\kappa$, which is performed using height-function method \cite{cummins2005estimating}. In this approach, the volume fraction is summed along grid lines to obtain the averaged interface position within each cell column (or row). For instance, in the $y$-direction, the height function is defined as,
\begin{equation}
  h_{i}=\sum_{j}c_{i,j}\Delta
\end{equation}
where the range of summation is adaptively chosen according to the local interface topology (see \cite{popinet2009accurate} for details). Once the height function is determined, the interface normal vector $\mathbf{n}$ and curvature $\kappa$ in the interfacial cell $(i,j)$ can be calculated as
\begin{equation}
  \mathbf{n}=(-h_x,\;1)
  \label{normal}
\end{equation}
\begin{equation}
  \kappa=\frac{h_{xx}}{(1+h_x^2)^{3/2}}
  \label{curvature}
\end{equation}
Here, $h_x$ and $h_{xx}$ denote the first and second derivatives of the height function, respectively, which are evaluated using central finite differences, i.e., $h_x=(h_{i+1,j}-h_{i-1,j})/2\Delta$ and $h_{xx}=(h_{i+1,j}-2h_{i,j}+h_{i-1,j})/\Delta^2$. An alternative approach available in \textit{Basilisk} for estimating the interface normal is the Mixed-Youngs-Centred (MYC) method \cite{aulisa2007interface}.

\subsection{Contact angle enforcement}

In this section, we first review existing height-function based approaches for enforcing contact angles on flat solid boundaries. We then propose an improved scheme for handling extreme contact angles (i.e., angles close to 0° or 180°), which achieves higher accuracy than conventional formulations. Finally, the method is further generalized to impose arbitrary contact angles on solid surfaces with complex geometries.

\subsubsection{Existing height-function based contact angle model}
\label{sec:cae}

\begin{figure}[tbp]
    \centering
    \begin{subfigure}[b]{0.4\textwidth}
        \centering
        \includegraphics[width=\textwidth]{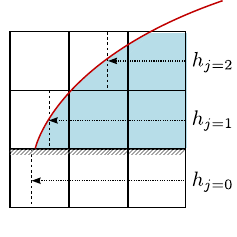}
        \caption{}
        \label{fig:sub0-2-1}
    \end{subfigure}
    \hspace{2cm}
    \begin{subfigure}[b]{0.4\textwidth}
        \centering
        \includegraphics[width=\textwidth]{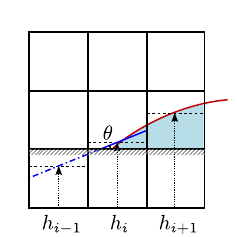}
        \caption{}
        \label{fig:sub0-2-2}
    \end{subfigure}
    \caption{Schematic of existing height-function based contact angle models proposed by Afkhami and Bussmann \cite{afkhami2008height}. (a) Horizontal height function. (b) Vertical height function. In (b), a linear interface (blue solid line) forming an angle $\theta$ with the solid surface is reconstructed in the contact line cell and extended to the neighboring column (blue dash-dotted line) to define the height function $h_{i-1}$.}
    \label{fig0-2}
\end{figure}

Afkhami and Bussmann \cite{afkhami2008height} proposed a height-function based scheme for enforcing contact angles in two-dimensional VOF simulations. For moderate contact angles ($45^\circ<\theta<135^\circ$), the height function defined parallel to the solid surface (hereafter referred to as horizontal height function) can be used to impose the contact angle boundary condition. Specifically, as illustrated in Fig.~\ref{fig0-2}(a), the height function in the ghost cell layer, $h_{j=0}$, is assigned from the adjacent height function $h_{j=1}$ and the prescribed contact angle $\theta$ as
\begin{equation}
  h_{j=0} = h_{j=1} + \Delta / \tan\theta,
\end{equation}
where $\Delta$ denotes the grid size.

For small or large contact angles ($\theta<45^\circ$ or $\theta>135^\circ$), the height function defined perpendicular to the solid surface (hereafter referred to as vertical height function) is employed instead. As illustrated in Fig.~\ref{fig0-2}(b), a linear interface segment forming an angle $\theta$ with the solid surface is first reconstructed within the contact line cell (i.e., the interfacial cell containing contact line) and then extended into the neighboring non-interfacial cell column. From the equation of this reconstructed interface, the height functions $h_{i}$ and $h_{i-1}$ can be determined, which satisfies the following relation
\begin{equation}
  h_{i-1} = h_{i} - \Delta \cdot \tan\theta.
\end{equation}
Once the height functions neighboring the contact-line cell are defined, the interface normal vector and curvature in that cell can be directly evaluated using Eqs.~(\ref{normal}) and (\ref{curvature}).

In $Basilisk$, only the contact angle model based on the horizontal height function formulation is available. Consequently, it cannot accurately handle extreme contact angles (e.g., $\theta < 30^\circ$ or $\theta > 150^\circ$). Although the vertical height function can be used as an alternative in such cases, its accuracy is inferior to that of the horizontal formulation. This is because the height function in the contact-line cell is derived from the position of the reconstructed linear interface, which itself contains a first-order error. The extrapolated height function $h_{i-1}$, obtained by extending the reconstructed interface in the contact-line cell (see Fig.~\ref{fig0-2}(b)), therefore inherits this first-order error. As will be demonstrated in Sec.~\ref{subsec:stds}, enforcing the contact angle via the vertical height function is less accurate than using the horizontal one.

For the application of extreme contact angles, the use of the vertical height function is indispensable. However, as discussed above, this approach suffers from reduced accuracy due to the definition of the height function in the contact line cell and particularly its extrapolation into the neighboring non-interfacial cell. To address this limitation, we develop an improved model for enforcing extreme contact angles, which still relies on the vertical height function but achieves higher accuracy.

\begin{figure}[t]%% placement specifier
    \centering%% For centre alignment of image.
    \includegraphics[width=0.4\textwidth]{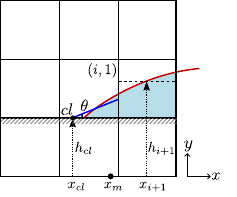}
    \caption{Schematic of the improved model for enforcing extreme contact angles. The height function $h_{cl}$ denotes the interface height at the contact line position $cl$, which equals $\Delta$ when measured relative to the bottom boundary of cell $(i,0)$. $x_m$ is the midpoint between $x_{cl}$ and $x_{i+1}$.}\label{fig0-3}
\end{figure}

\subsubsection{Improved contact angle enforcement for extreme angles}
\label{sec:icae}

The new model is illustrated in Fig.~\ref{fig0-3}. Similar to the method of Afkhami and Bussmann \cite{afkhami2008height} described in Sec.~\ref{sec:cae}, a linear interface segment forming an angle $\theta$ with the solid surface is first reconstructed in the contact-line cell. From its equation, a height function $h_{i}$ similar to that shown in Fig.~\ref{fig0-2}(b) can be obtained, which is used solely for estimating the interface normal vector and curvature in the adjacent interfacial cell $(i+1,1)$. For the contact-line cell $(i,1)$ itself, a different procedure is adopted. Specifically, the intersection point $cl$ between the reconstructed interface and the solid surface is identified and regarded as the contact-line position. At this point, an accurate height function $h_{cl}$ can be defined to represent the interface height at $x_{cl}$. With $h_{cl}$, $h_{i+1}$, and the prescribed contact angle $\theta$, the first and second derivatives of the height function in the contact-line cell can then be evaluated as
\begin{equation}\label{hxcl}
  h_{x,cl} = \tan\theta
\end{equation}
\begin{equation}\label{hxm}
  h_{x,m} = \frac{h_{i+1} - h_{cl}}{x_{i+1}-x_{cl}}
\end{equation}
\begin{equation}\label{hxxcl}
  h_{xx,cl} = \frac{2(h_{x,m} - h_{x,cl})}{x_{i+1}-x_{cl}},
\end{equation}
where $h_{x,m}$ denotes the first derivative of the height function evaluated at the midpoint between $x_{i+1}$ and $x_{cl}$. With $h_{x,cl}$ and $h_{xx,cl}$ thus defined, the curvature in the contact line cell can be calculated using Eq.~(\ref{curvature}).

In the proposed model, the key element is the contact-line position $cl$, which is obtained from the reconstructed linear interface in the contact-line cell.  To ensure a smooth variation of the estimated curvature in the contact line cell, the contact line position $cl$ is required to move continuously along the solid surface during interface advection. This requirement is fulfilled by  constraining the interface normal vector in the contact-line cell to
\begin{equation}
\mathbf{n}_{cl} = (-\cos\theta,\;\sin\theta),
\end{equation}
where $\theta$ is the prescribed contact angle. With a continuously varying volume fraction $c$ and a fixed normal vector $\mathbf{n}_{cl}$ in the contact-line cell, the resulting contact-line position $cl$ moves smoothly on the solid surface, which is crucial for the accuracy of the present model.

\subsubsection{Extension to complex geometries}
\label{sec:ecg}

The improved contact-angle model presented in Sec.\ref{sec:icae} was originally developed for extreme contact angles on flat solid surfaces. Here, we show that the same formulation can be naturally extended to impose arbitrary contact angles on solid surfaces with complex geometries. Within the geometric VOF framework, existing contact angle models for complex geometries have typically relied on ghost-cell extrapolation. For example, Tavares et al. \cite{tavares2024coupled} and Chen et al. \cite{chen2025volume} extrapolated the reconstructed linear interface, oriented at the prescribed angle, into the solid to assign ghost-cell volume fractions. However, such linear extrapolation performs poorly at extreme contact angles, although Chen et al. \cite{chen2025volume} partially alleviated this limitation by identifying the contact line cell prior to extrapolation. More recently, Huang et al. \cite{huang20252d} introduced a parabola-fitting scheme to approximate the height-function distribution near the contact line, which improves accuracy compared with linear extrapolation but involves a more complex implementation. In contrast, the present approach can be regarded as a direct extension of the flat-wall model in Sec.\ref{sec:icae}. It can enforce arbitrary contact angles on complex geometries with accuracy comparable to the parabola-fitting method, while retaining a much simpler implementation.

\begin{figure}[t]%% placement specifier
    \centering%% For centre alignment of image.
     \includegraphics[width=0.4\textwidth]{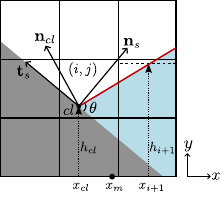}
    \caption{Schematic of the contact angle model on an embedded solid surface. The height function $h_{cl}$ represents the interface height at the contact line position $cl$. $\mathbf{n}_{cl}$ is the interface normal in the contact line cell $(i,j)$. $\mathbf{n}_{s}$ and $\mathbf{t}_{s}$ denote the normal and tangential vectors of the solid boundary within the contact line cell, respectively.}\label{fig0-6}
\end{figure}

Taking the height function defined along the $y$-direction as an example, and as illustrated in Fig.~\ref{fig0-6}, a linear interface segment forming the prescribed contact angle $\theta$ with the solid boundary is first reconstructed in the contact-line cell $(i,j)$. Based on the equation of this reconstructed interface, a height function $h_i$ is defined at the centerline of the cell column (see Fig.~\ref{fig0-2}(b)), and is used exclusively to estimate the interface normal vector and curvature in the adjacent interfacial column $(i+1,j)$. For the contact-line cell itself, the intersection point $cl$ between the reconstructed interface and the solid boundary is identified as the contact-line position. At this point, a height function $h_{cl}$ is defined, and the same formulations \eqref{hxcl}, \eqref{hxm}, and \eqref{hxxcl} are then applied to evaluate the interface curvature in cell $(i,j)$.

Compared with the flat-surface model in Sec.\ref{sec:icae}, the only difference lies in the interface reconstruction within the contact-line cell. Here, the interface normal $\mathbf{n}_{cl}$ is determined jointly by the prescribed contact angle $\theta$ and the local orientation of the embedded solid boundary, namely,
\begin{equation} \label{ncl2}
\mathbf{n}_{cl} = \mathbf{n}_{s}\cos\theta + \mathbf{t}_{s}\sin\theta,
\end{equation}
where $\mathbf{n}_{s}$ and $\mathbf{t}_{s}$ denote the unit normal and tangential vectors of the solid boundary within the contact line cell, as shown in Fig.~\ref{fig0-6}. The implementation details for computing $\mathbf{n}_{cl}$ are available in the $Basilisk$ sandbox \cite{popinetBasilisk}.

As in the case of flat solid surfaces, the contact-line position $cl$ is expected to move smoothly along the embedded solid boundary during interface advection, since the curvature estimation in the contact line cell directly depends on the contact line position. To satisfy this requirement, the normal vector of the reconstructed interface in the contact-line cell is determined by Eq.\eqref{ncl2} rather than evaluated from Eq.\eqref{normal}. Combined with a continuous variation of the volume fraction in the contact-line cell, this treatment ensures smooth advection of the contact line along the embedded solid surface.

Finally, it is important to identify the contact-line cell before enforcing the contact angle boundary condition. In this study, we adopt a procedure similar to that proposed by Afkhami and Bussmann \cite{afkhami2008height}. For contact angles smaller than $90^{\circ}$, the three-phase cell ($0 < c < c_s < 1$) neighboring a pure-gas cut cell ($c = 0$ and $0 < c_s < 1$) is identified as the contact-line cell. Conversely, for contact angles larger than $90^{\circ}$, the three-phase cell neighboring a pure-liquid cut cell ($c > 0$ and $c = c_s$) is identified as the contact-line cell. This criterion ensures a consistent and robust identification of the contact-line cell on embedded solid boundaries.

%% Use \subsubsection, \paragraph, \subparagraph commands to 
%% start 3rd, 4th and 5th level sections.
%% Refer following link for more details.
%% https://en.wikibooks.org/wiki/LaTeX/Document_Structure#Sectioning_commands

\section{Results}

In this section, several test cases are conducted to validate the numerical schemes proposed in Sec.~\ref{sec2}. First, the pure advection of a circular interface around a cylinder is examined to evaluate the performance of the interface reconstruction and advection algorithms developed for cut cells. Second, the surface-tension-driven spreading of a droplet on a flat solid surface is simulated to demonstrate the accuracy and effectiveness of the improved contact-angle model introduced in Sec.\ref{sec:icae} for handling extreme contact angles. Third, the enforcement of contact angles on embedded solid boundaries is assessed through sessile droplet spreading on both flat and curved embedded surfaces, as well as droplet penetration in porous media. Finally, two micro-channel flow configurations are considered to reveal the remaining limitations of the present contact-angle enforcement approach for embedded solid boundaries.

\subsection{Circular interface advection around a cylinder}

\begin{figure}[t]
    \centering
    \begin{subfigure}[b]{0.45\textwidth}
        \centering
        \includegraphics[width=\textwidth]{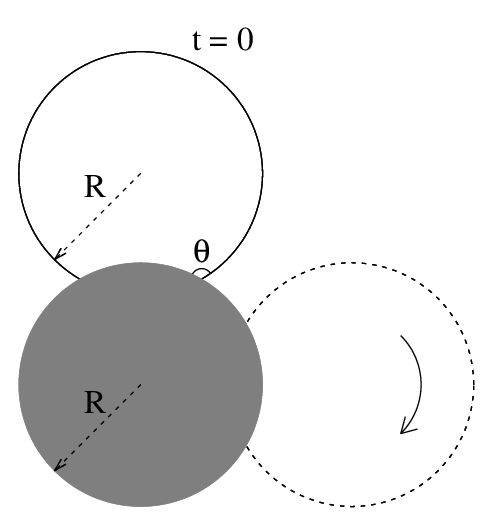}
        \caption{}
        \label{fig:sub2-1-1}
    \end{subfigure}
    \hfill
    \begin{subfigure}[b]{0.45\textwidth}
        \centering
        \includegraphics[width=\textwidth]{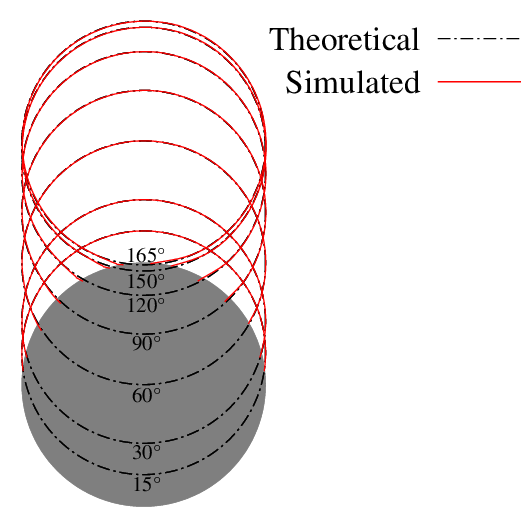}
        \caption{}
        \label{fig:sub2-1-2}
    \end{subfigure}
    \caption{Interface advection test. (a) Clockwise rotation of a circular interface of radius $R$ around a cylinder of the same radius. Solid and dashed arcs indicate the initial and intermediate interface positions, respectively. (b) Interface shapes after one full rotation for $\theta = 15^{\circ}, 30^{\circ}, 60^{\circ}, 90^{\circ}, 120^{\circ}, 150^{\circ}$, and $165^{\circ}$. The black dash-dotted lines represent the theoretical circular profiles. The grid resolution is 18 cells per radius $R$.}
    \label{fig2-1}
\end{figure}

In this subsection, we validate the interface reconstruction and advection algorithms developed for cut cells in Sec.~\ref{sec:iftrack}. As illustrated in Fig.~\ref{fig2-1}(a), a pure advection test is performed in which a circular fluid–fluid interface of radius $R$ is transported around a stationary cylinder of the same radius by the prescribed velocity field
\begin{equation}
\left\{
\begin{aligned}
u_x &= \;\;\;2\pi y, \\
u_y &= -2\pi x.
\end{aligned}
\right.
\end{equation}
with the coordinate origin located at the cylinder center. This velocity field induces a clockwise rotation of the interface around the cylinder. To satisfy the prescribed contact angle $\theta$, the center of the circular interface is initially positioned at a distance $d=\sqrt{2(1-\cos\theta)}R$ from the cylinder center. Unless otherwise specified, the grid resolution is set to 18 cells per radius $R$.

The interface shape after one full rotation around the cylinder is shown in Fig.~\ref{fig2-1}(b). The initial circular interface profiles are also plotted as theoretical references for comparison. From $15^{\circ}$ to $165^{\circ}$, the final interface profiles agree well with the initial ones, with noticeable differences appearing only near the contact line at large contact angles. This discrepancy arises because the interface normal vector in the contact-line cell is fixed to always form the prescribed angle $\theta$ with the solid boundary, which introduces errors in the interface advection near the contact line.

\begin{figure}[t]%% placement specifier
    \centering%% For centre alignment of image.
    \includegraphics[width=0.6\textwidth]{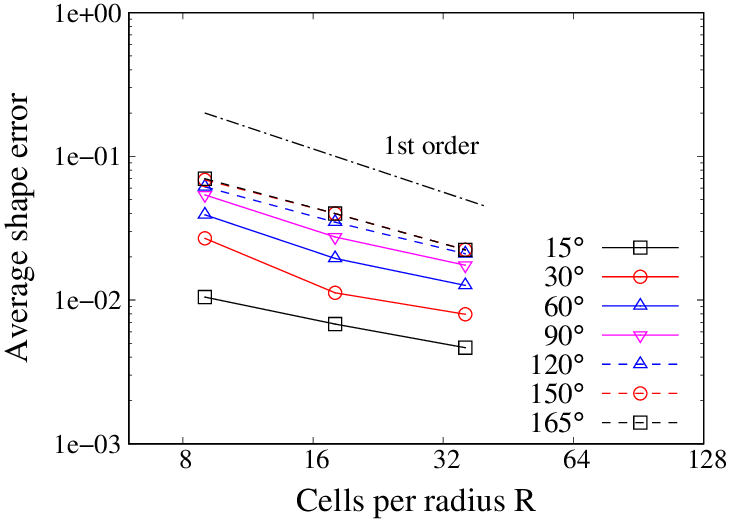}
    \caption{Average interface shape error of the advected circular interface after one full rotation around the cylinder for different contact angles.}\label{fig2-2}
\end{figure}

To further quantify the difference between the final and initial interface shapes shown in Fig.~\ref{fig2-1}(b), the average error of the computed volume fraction is evaluated over all interfacial cells as
\begin{equation}\label{eq:ec}
	E_c =
	\sqrt{
		\frac{1}{N_i}
		\sum_{i=1}^{N_i}
		\left( c_i - c_{i,\text{exact}} \right)^2
	},
\end{equation}
where $N_i$ denotes the total number of interfacial cells, $c_i$ is the computed volume fraction, and $c_{i,\text{exact}}$ is the corresponding exact value. As shown in Fig.~\ref{fig2-2}, the shape error decreases monotonically with grid refinement for all contact angles, exhibiting nearly first-order convergence. This behavior is consistent with the linear interface reconstruction employed in the present method and confirms the accuracy of the interface reconstruction and advection algorithms designed for cut cells.

\subsection{Surface-tension driven spreading on a flat solid surface}
\label{subsec:stds}

We consider a semicircular droplet of radius $R = 0.5\,\text{m}$ initially placed on a flat solid surface, on which a prescribed contact angle $\theta \neq 90^{\circ}$ is enforced. In the absence of gravity, the droplet spreads or retracts until reaching an equilibrium circular shape consistent with the imposed contact angle. The droplet and the surrounding fluid are assumed to have identical density and dynamic viscosity, $\rho = 1\,\text{kg}/\text{m}^3$ and $\mu = 0.04\,\text{kg}/\text{m}\cdot \text{s}$, respectively, while the surface-tension coefficient between the two fluids is $\sigma = 0.1\,\text{N}/\text{m}$. The corresponding dimensionless Ohnesorge number is $\mathrm{Oh} = \mu/\sqrt{\rho \sigma R} = 0.179$. A relatively large Ohnesorge number is chosen here to suppress oscillations of the droplet around the equilibrium state and to accelerate its relaxation toward equilibrium. Unless explicitly stated otherwise, the grid resolution is set to 16 cells per initial droplet radius.

\begin{figure}[htbp]
	\centering
	\begin{subfigure}[b]{\textwidth}
		\centering
		\includegraphics[width=0.99\textwidth]{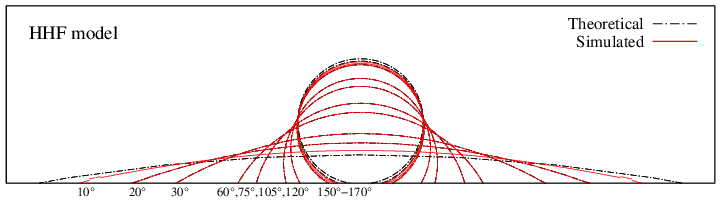}
		\caption{}
		\label{fig:sub1-1-1}
	\end{subfigure}
	\begin{subfigure}[b]{\textwidth}
		\centering
		\includegraphics[width=0.99\textwidth]{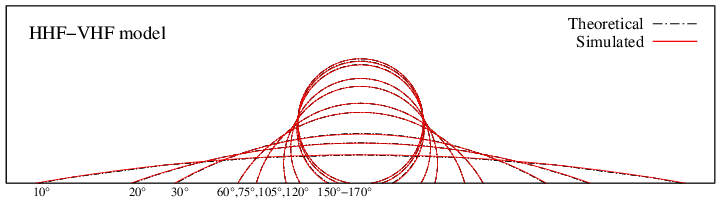}
		\caption{}
		\label{fig:sub1-1-2}
	\end{subfigure}
	\begin{subfigure}[b]{\textwidth}
		\centering
		\includegraphics[width=0.99\textwidth]{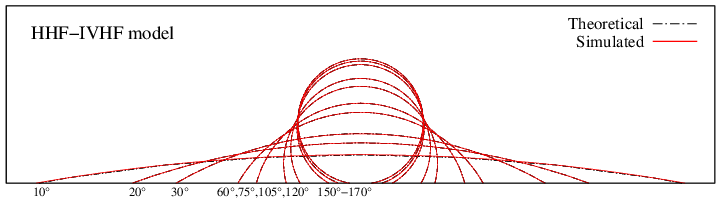}
		\caption{}
		\label{fig:sub1-1-3}
	\end{subfigure}
	\caption{Sessile droplet spreading at contact angles of $10^{\circ}$, $20^{\circ}$, $30^{\circ}$, $60^{\circ}$, $75^{\circ}$, $105^{\circ}$, $120^{\circ}$, $150^{\circ}$, $160^{\circ}$, and $170^{\circ}$. Contact angles are enforced using (a) the existing horizontal HF (HHF) model only \cite{afkhami2008height}, (b) the existing horizontal and vertical HF (HHF--VHF) model \cite{afkhami2008height}, and (c) the existing horizontal \cite{afkhami2008height} and the improved vertical HF (HHF--IVHF) model proposed in Sec.~\ref{sec:icae}. Grid resolution is 16 cells per initial droplet radius.}
	\label{fig1-1}
\end{figure}

Fig.~\ref{fig1-1}(a) displays the equilibrium droplet shapes obtained using the horizontal height function (HF) model \cite{afkhami2008height}, described in Sec.~\ref{sec:cae} and hereafter referred to as the HHF model. The droplet is regarded as reaching equilibrium when the maximum change in the volume fraction between two successive time steps in any cell of the computational domain falls below $10^{-6}$, or when the simulation time exceeds $4t_{\mu}$, where $t_{\mu} = D^2/\mu$ is the viscous time scale with $D = 2R$. For contact angles between $30^{\circ}$ and $150^{\circ}$, the simulated droplet profiles agree well with the theoretical solutions (ideal circular caps). This range also corresponds to the recommended validity range of the horizontal HF model in $Basilisk$. However, at very small or very large contact angles, the equilibrium interface shape near the contact line becomes discontinuous, demonstrating the poor performance of the horizontal HF model under these extreme contact angles.

Fig.~\ref{fig1-1}(b) presents the equilibrium droplet shapes obtained using the combined horizontal and vertical HF models (see Sec.~\ref{sec:cae}) proposed by Afkhami et al. \cite{afkhami2008height}, hereafter referred to as the HHF-VHF model. Compared with Fig.~\ref{fig1-1}(a), noticeable differences appear only at extreme contact angles, which are now imposed by the vertical HF model. As expected, for both very small contact angles (down to $10^{\circ}$) and very large contact angles (up to $170^{\circ}$), the simulated equilibrium droplet profiles agree well with the theoretical solutions, demonstrating the effectiveness of the vertical HF model in treating extreme contact angles.

\begin{figure}[tbp]
	\centering
	\begin{subfigure}[b]{0.49\textwidth}
		\centering
		\includegraphics[width=\textwidth]{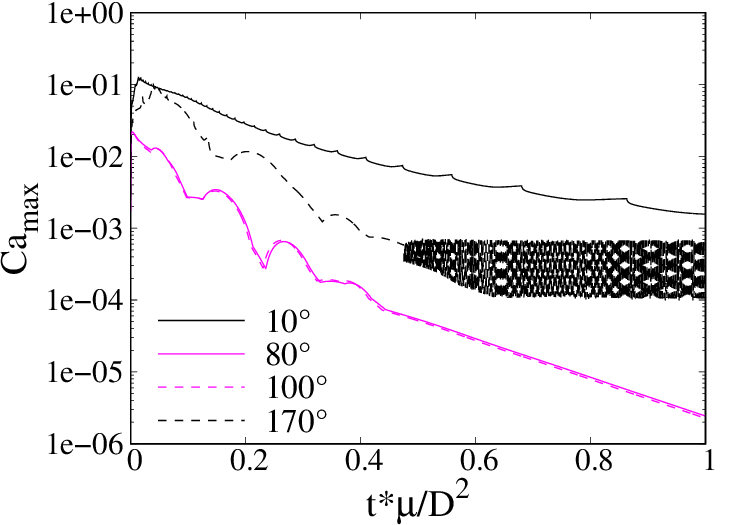}
		\caption{}
		\label{fig:sub1-2-1}
	\end{subfigure}
	\hfill
	\begin{subfigure}[b]{0.49\textwidth}
		\centering
		\includegraphics[width=\textwidth]{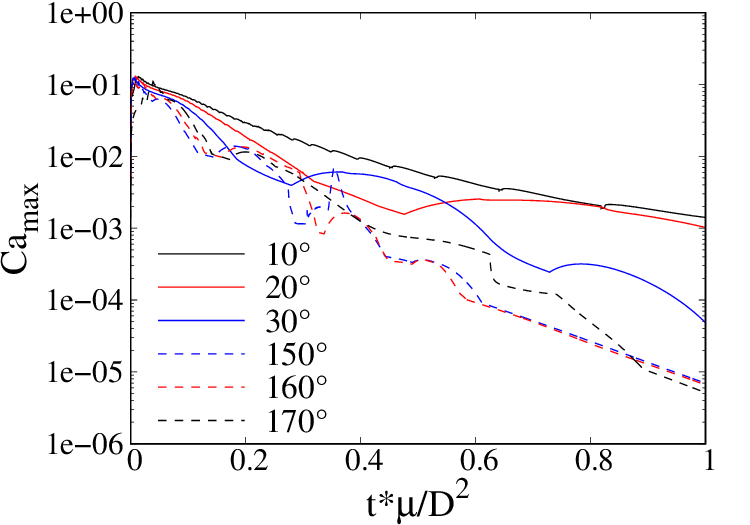}
		\caption{}
		\label{fig:sub1-2-2}
	\end{subfigure}
	\caption{Time evolution of the maximum velocity in the computational domain, defined as $\mathrm{Ca}_{\max} = \mu |u_{\max}|/\sigma$, for sessile droplet spreading on a flat solid surface. Contact angles are enforced using (a) the existing horizontal and vertical HF models \cite{afkhami2008height}, and (b) the improved vertical HF model proposed in Sec.~\ref{sec:icae}.}
	\label{fig1-2}
\end{figure}

To further compare the performance of the horizontal and vertical HF models in enforcing contact angles, we plot the time evolution of the maximum velocity in the computational domain, defined as $\mathrm{Ca}_{\max} = \mu |u_{\max}|/\sigma$, as shown in Fig.~\ref{fig1-2}(a). Four representative contact angles are selected: $80^{\circ}$ and $100^{\circ}$ for the horizontal HF model, and $10^{\circ}$ and $170^{\circ}$ for the vertical HF model, under which both models are expected to perform accurately. The results, however, differ significantly between these two models. For the horizontal HF model, the maximum velocity decreases continuously and reaches $10^{-6}$ by $t_{\mu}$. In contrast, for the vertical HF model, the velocity decay is very slow at $10^{\circ}$, likely due to the large spreading extent in this case, and exhibits oscillations in the final equilibrium state at $170^{\circ}$. This behavior stems from curvature errors at the contact line, caused by the inaccurate definition of the height function in the contact-line cell and its extrapolation into the neighboring non-interfacial cell column, as discussed in Sec.\ref{sec:cae}.

\begin{figure}[htbp]
	\centering
	\begin{subfigure}[b]{0.49\textwidth}
		\centering
		\includegraphics[width=\textwidth]{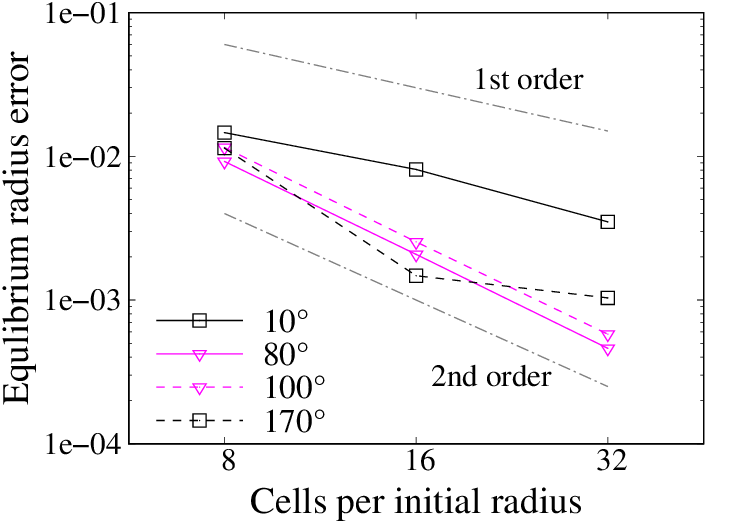}
		\caption{}
		\label{fig:sub1-3-1}
	\end{subfigure}
	\hfill
	\begin{subfigure}[b]{0.49\textwidth}
		\centering
		\includegraphics[width=\textwidth]{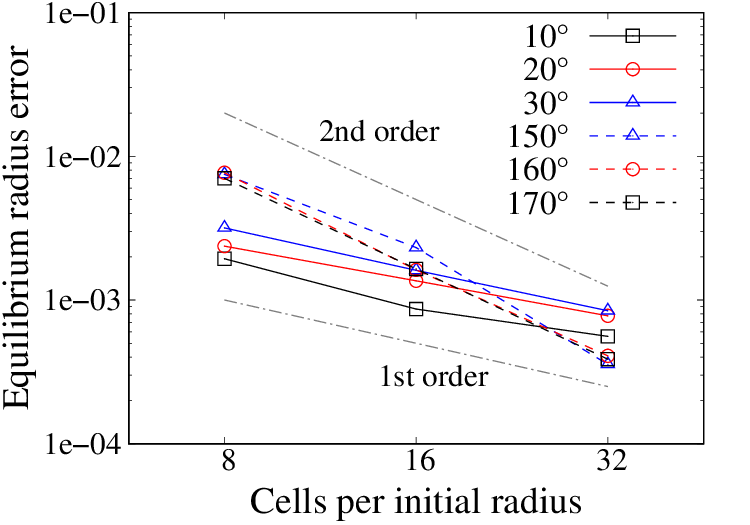}
		\caption{}
		\label{fig:sub1-3-2}
	\end{subfigure}
		\begin{subfigure}[b]{0.49\textwidth}
		\centering
		\includegraphics[width=\textwidth]{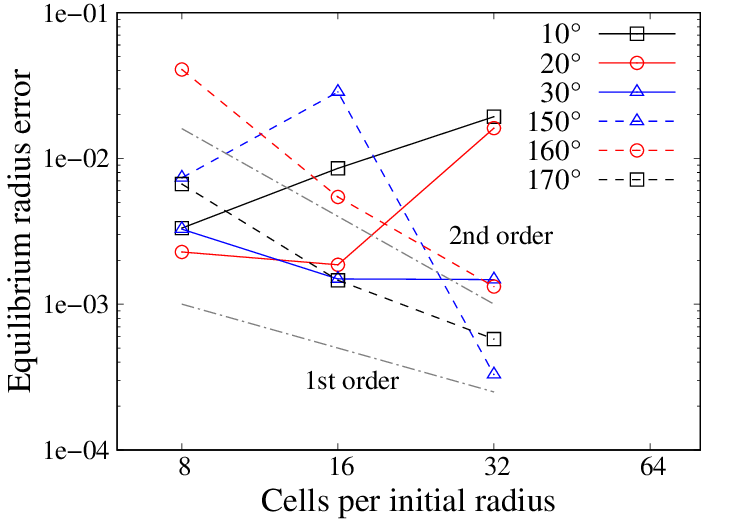}
		\caption{}
		\label{fig:sub1-3-3}
	\end{subfigure}
	\caption{Relative error of the equilibrium droplet radius as a function of grid resolution for sessile droplet spreading on a flat solid surface. Contact angles are enforced using (a) the existing horizontal and vertical HF models \cite{afkhami2008height}, and (b,c) the improved vertical HF model proposed in Sec.~\ref{sec:icae}. In panels (a) and (b), the equilibrium radius error is evaluated over all interfacial cells, whereas in panel (c) it is evaluated using only the first cell layer adjacent to the solid wall.}
	\label{fig1-3}
\end{figure}

Fig.~\ref{fig1-3}(a) presents the relative error of the averaged equilibrium droplet radius $R_e$, evaluated over the entire interface and defined as
\begin{equation}\label{eq:Er}
	E_R =
		 \frac{1}{N_r}
		\sum_{k=1}^{N_r} \left(\frac{1}{R_\text{exact} } 
		\left|\frac{1}{N_i} \sum_{j=1}^{N_i} R_j^{(k)} - R_\text{exact}\right|\right),
\end{equation}
where $N_i$ denotes the total number of interfacial cells, $R_{j}^{(k)}$ is the local radius computed at the $j$-th interfacial cell for the $k$-th droplet size, and $R_{\text{exact}}$ is the theoretical equilibrium radius given by
\begin{equation} \label{Rthe1}
R_{\text{exact}} = \sqrt{\frac{0.5\pi R^2}{\theta - \sin\theta\cos\theta}}.
\end{equation}
The average over $N_r$ in Eq.~\eqref{eq:Er} is computed from $N_r=100$ droplets with different sizes, whose initial radii are uniformly distributed in $[R-\Delta,\,R+\Delta]$, where $\Delta$ is the grid size. This sampling ensures that a wide range of possible interface–grid intersection configurations is covered. For the horizontal HF model, the error in $R_e$ decreases with grid refinement at a second-order rate, consistent with the known accuracy of curvature estimation by the height-function method \cite{cummins2005estimating}. In contrast, for the vertical HF model, the convergence rate drops to first order at $\theta = 10^{\circ}$ and deteriorates further at $\theta = 170^{\circ}$, where the error decreases initially but eventually saturates upon further refinement. This poor convergence occurs because the curvature estimation by the vertical HF does not improve with grid refinement near the contact line. At $\theta = 10^{\circ}$, the impact of this local error is partially mitigated by the large spreading extent of the droplet, whereas at $\theta = 170^{\circ}$ it becomes dominant, also giving rise to significant spurious flows near the contact line, as verified in Fig.~\ref{fig1-2}(a). Combining the results in Fig.~\ref{fig1-2}(a) and Fig.~\ref{fig1-3}(a), we conclude that the accuracy of the vertical HF model in enforcing contact angles is inferior to that of the horizontal HF model.

Finally, we assess the performance of the improved vertical HF model for contact angle enforcement proposed in Sec.~\ref{sec:icae}, hereafter referred to as the IVHF model. Fig.~\ref{fig1-1}(c) shows the equilibrium droplet shapes at different contact angles. For both very small and very large contact angles, which are enforced using the IVHF model, the resulting equilibrium profiles agree closely with the theoretical circular-cap shapes. To further quantify this behavior, the relative errors of the equilibrium droplet radius are compared across three models: the horizontal HF (HHF) model, the vertical HF (VHF) model, and the IVHF model. The HHF and VHF models are described in Sec.~\ref{sec:cae}. The results are summarized in Table~\ref{tab1-3-1}(a), where only contact angles below $30^{\circ}$ and above $150^{\circ}$ are reported.

In this range of contact angles, the vertical HF model is generally expected to achieve higher accuracy than the horizontal HF model. However, the results in Table~\ref{tab1-3-1}(a) show that the VHF model yields larger errors than the HHF model at all considered angles except $10^{\circ}$ and $170^{\circ}$. Only at $10^{\circ}$ does the VHF model produce noticeably smaller errors than the HHF model. This outcome appears counterintuitive when compared with the droplet equilibrium shapes shown in Figs.~\ref{fig1-1}(a) and \ref{fig1-1}(b). A plausible explanation lies in the error definition: the equilibrium radius error is averaged over the entire interface, which tends to mask the local curvature inaccuracy near the contact line. In contrast, the IVHF model consistently outperforms the VHF model across all tested contact angles, except at $160^{\circ}$. Even compared with the HHF model, the IVHF model yields significantly smaller errors at small contact angles and comparable errors at large contact angles, thereby demonstrating its superior accuracy relative to VHF.

\begin{table}[t]
\centering%% For centre alignment of tabular.
\caption{Errors of the equilibrium droplet shape at contact angles below $30^{\circ}$ and above $150^{\circ}$. Results are shown for three contact-angle enforcement models: the existing horizontal HF (HHF) model  \cite{afkhami2008height}, the existing vertical HF (VHF) model \cite{afkhami2008height}, and the improved vertical HF (IVHF) model proposed in Sec.~\ref{sec:icae}. (a) Relative error of the equilibrium droplet radius. (b) Shape error of the equilibrium droplet. Bold numbers indicate the smallest error in each column.} 
\label{tab1-3-1}
\begin{subtable}{\textwidth}
	\centering
	\caption{Relative errors of the equilibrium droplet radius}
\begin{tabular}{l c c c c c c}%% Table column specifiers
%% Tabular cells are separated by &
    \toprule
       & $10^{\circ}$ & $20^{\circ}$ & $30^{\circ}$ & $150^{\circ}$ & $160^{\circ}$ & $170^{\circ}$\\ %% A tabular row ends with \\
    \midrule
    \textbf{HHF} & $6.31 \mathrm{e}{-2}$ & $1.22 \mathrm{e}{-3}$ & $3.16 \mathrm{e}{-3}$ & $\mathbf{3.37 {e}{-3}}$ & $\mathbf{2.39 {e}{-3}}$ & $6.10 \mathrm{e}{-3}$ \\
    \textbf{VHF} & $5.56 \mathrm{e}{-3}$ & $1.87 \mathrm{e}{-2}$ & $2.06 \mathrm{e}{-2}$ & $9.12 \mathrm{e}{-3}$ & $7.54 \mathrm{e}{-3}$ & $5.60 \mathrm{e}{-3}$ \\
    \textbf{IVHF} & $\mathbf{7.96 {e}{-4}}$ & $\mathbf{7.39 {e}{-4}}$ & $\mathbf{2.26 {e}{-4}}$ & $4.00 \mathrm{e}{-3}$ & $9.45 \mathrm{e}{-3}$ & $\mathbf{4.71 {e}{-3}}$ \\
    \bottomrule
\end{tabular}
\end{subtable}

\vspace{0.6em}

\begin{subtable}{\textwidth}
	\centering
	\caption{Shape errors of the equilibrium droplet}
	\begin{tabular}{l c c c c c c}%% Table column specifiers
		%% Tabular cells are separated by &
		\toprule
		& $10^{\circ}$ & $20^{\circ}$ & $30^{\circ}$ & $150^{\circ}$ & $160^{\circ}$ & $170^{\circ}$\\ %% A tabular row ends with \\
		\midrule
		\textbf{HHF} & $6.62 \mathrm{e}{-2}$ & $1.43 \mathrm{e}{-3}$ & $7.22 \mathrm{e}{-4}$ & $3.28 \mathrm{e}{-3}$ & $8.12 \mathrm{e}{-3}$ & $4.68 \mathrm{e}{-2}$ \\
		\textbf{VHF} & $\mathbf{5.91 {e}{-4}}$ & $4.05 \mathrm{e}{-3}$ & $5.17 \mathrm{e}{-3}$ & $5.99 \mathrm{e}{-3}$ & $\mathbf{2.99 {e}{-3}}$ & $\mathbf{5.99 {e}{-4}}$ \\
		\textbf{IVHF} & $6.17 \mathrm{e}{-4}$ & $\mathbf{5.41 {e}{-4}}$ & $\mathbf{5.74 {e}{-4}}$ & $\mathbf{6.82 {e}{-4}}$ & $3.65 \mathrm{e}{-3}$ & $1.55 \mathrm{e}{-3}$ \\
		\bottomrule
	\end{tabular}
\end{subtable}
\end{table}

To further clarify the apparent inconsistency between the results in Fig.~\ref{fig1-1} and Table~\ref{tab1-3-1}(a), we additionally evaluate the shape error at equilibrium for the three models. The shape error is defined as the root-mean-square difference between the computed volume-fraction field and that of the corresponding ideal circular cap, as given by Eq.~\eqref{eq:ec}, but evaluated over all cells in the computational domain. The results are summarized in Table~\ref{tab1-3-1}(b). For small contact angles, the VHF model yields a substantially smaller error than the HHF model at $10^{\circ}$ but larger errors at $20^{\circ}$ and $30^{\circ}$. In contrast, the IVHF model  consistently produces small errors across all three angles, outperforming both HHF and VHF. For large contact angles, the VHF model performs better than the HHF model at $160^{\circ}$ and $170^{\circ}$, while yielding a larger error at $150^{\circ}$. The IVHF model remains more accurate than the HHF model across all three large-angle cases. However, it produces a noticeably larger error at $170^{\circ}$ relative to VHF.

The temporal evolution of the maximum velocity and the convergence behavior of the equilibrium radius error are examined for the improved vertical HF model. In Fig.~\ref{fig1-2}(b), the maximum velocity in the computational domain decays continuously with time for all extreme contact angles. The spurious flows observed with the conventional vertical HF model---manifested as oscillatory velocities at large contact angles [cf. Fig.\ref{fig1-2}(a)]---are effectively eliminated, demonstrating the improved accuracy of the present model. Fig.~\ref{fig1-3}(b) illustrates the grid convergence of the equilibrium radius error, which is evaluated based on Eq.~\eqref{eq:Er}. The error decreases with grid refinement at second-order accuracy for large contact angles and at first-order accuracy for small contact angles, which outperforms the results obtained with existing vertical HF model in Fig.~\ref{fig1-3}(a).

%\begin{figure}[t]%% placement specifier
%	\centering%% For centre alignment of image.
%	\includegraphics[width=0.55\textwidth]{figs/a11.eps}
%	\caption{Relative error of the equilibrium droplet radius evaluated in the first cell layer next to the solid wall. Contact angles are imposed using the improved vertical HF model.}\label{fig-a11}
%\end{figure}

To further elucidate the asymmetric convergence rates shown in Fig.~\ref{fig1-3}(b) for small and large contact angles, the relative errors of the equilibrium radius are re-evaluated using only the interfacial cells located in the first cell layer adjacent to the solid wall. The results are presented in Fig.~\ref{fig1-3}(c). It is important to note that the formulations \eqref{hxcl}--\eqref{hxxcl} employed to estimate the curvature in the contact-line cell do not guarantee a decreasing curvature error with grid refinement. As illustrated in Fig.~\ref{fig0-3}, although the height function $h_{cl}$ at the contact-line position is accurate by definition, its horizontal location $x_{cl}$ is identified from the reconstructed linear interface within the contact-line cell, which introduces a first-order error. Consequently, the curvature estimation---being directly related to the second derivative of the height function---may exhibit non-convergent behavior, as confirmed by the results for small contact angles in Fig.~\ref{fig1-3}(c). However, for large contact angles, the equilibrium radius errors within the first cell layer show unexpectedly good convergence, which explains the superior grid-convergence behavior observed in Fig.~\ref{fig1-3}(b).

\subsection{Surface-tension driven spreading on embedded solid surfaces}

In this subsection, we examine the contact angle enforcement on embedded solid boundaries [cf. Sec.~\ref{sec:ecg}] through surface-tension-driven droplet spreading. Unlike the test case in Sec.~\ref{subsec:stds}, where the solid boundary coincides with the domain boundary, here the solid surface is represented within the computational domain using the embedded boundary method. Two types of embedded solid surfaces are considered: an inclined flat surface and a circular surface. These two configurations have also been extensively investigated in the literature \cite{patel2017coupled,o2018volume,tavares2024coupled,huang20252d,chen2025volume}, which primarily focused on the equilibrium droplet shapes. In contrast, in addition to assessing the equilibrium droplet shapes, we also examine the temporal evolution of the velocity field within the computational domain.

\subsubsection{Inclined flat solid surface}
\label{subsubsec-3.3.1}

\begin{figure}[t]
	\centering
	\begin{subfigure}[b]{0.45\textwidth}
		\centering
		\includegraphics[width=\textwidth]{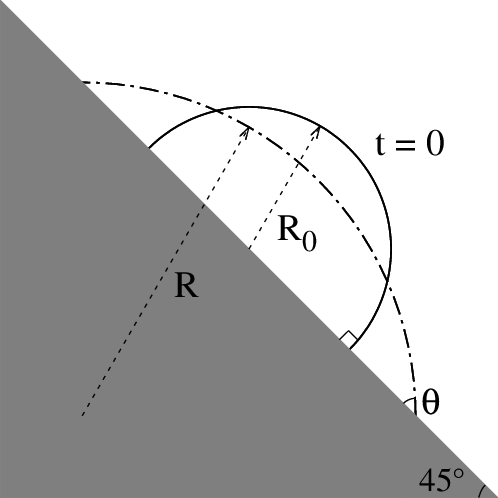}
		\caption{}
		\label{fig:sub3-1-1}
	\end{subfigure}
	\hfill
	\begin{subfigure}[b]{0.45\textwidth}
		\centering
		\includegraphics[width=\textwidth]{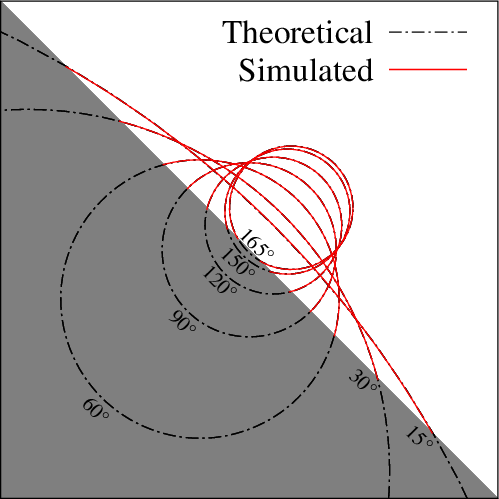}
		\caption{}
		\label{fig:sub3-1-2}
	\end{subfigure}
	\caption{Sessile droplet spreading on an inclined flat solid wall. (a) Schematic diagram. $R_0$ and $R$ denote the initial and equilibrium droplet radii, respectively. (b) Equilibrium droplet shapes at $\theta = 15^{\circ}$, $30^{\circ}$, $60^{\circ}$, $90^{\circ}$, $120^{\circ}$, $150^{\circ}$, and $165^{\circ}$. The black dash-dotted lines indicate the theoretical circular-cap shapes.}
	\label{fig3-1}
\end{figure}

As illustrated in Fig.~\ref{fig3-1}(a), we consider a semicircular droplet of radius $R_0 = 0.28\,\text{m}$ initially placed on an inclined flat solid surface. The surface is embedded within the computational domain and oriented at an angle of $45^{\circ}$ with respect to the grid lines. A prescribed contact angle $\theta \neq 90^{\circ}$ is enforced, driving the droplet to spread or retract until an equilibrium state is reached. Gravity is neglected, so the theoretical equilibrium shape corresponds to a circular cap with radius $R$, consistent with the imposed contact angle. The droplet and the surrounding fluid are assumed to have identical density and dynamic viscosity, $\rho = 1\,\text{kg}/\text{m}^3$ and $\mu = 0.04\,\text{kg}/(\text{m}\cdot \text{s})$, respectively, while the surface-tension coefficient is $\sigma = 0.1\,\text{N}/\text{m}$. The corresponding dimensionless Ohnesorge number is $\mathrm{Oh} = \mu/\sqrt{\rho \sigma R_0} = 0.239$. The grid resolution is set to 18 cells per initial droplet radius $R_0$.

\begin{figure}[t]
    \centering
    \begin{subfigure}[b]{0.49\textwidth}
        \centering
        \includegraphics[width=\textwidth]{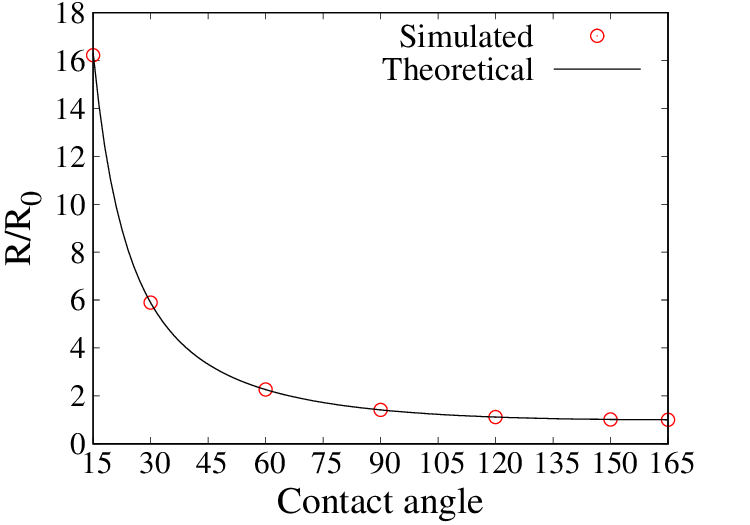}
        \caption{}
        \label{fig:sub3-2-1}
    \end{subfigure}
    \hfill
    \begin{subfigure}[b]{0.49\textwidth}
        \centering
        \includegraphics[width=\textwidth]{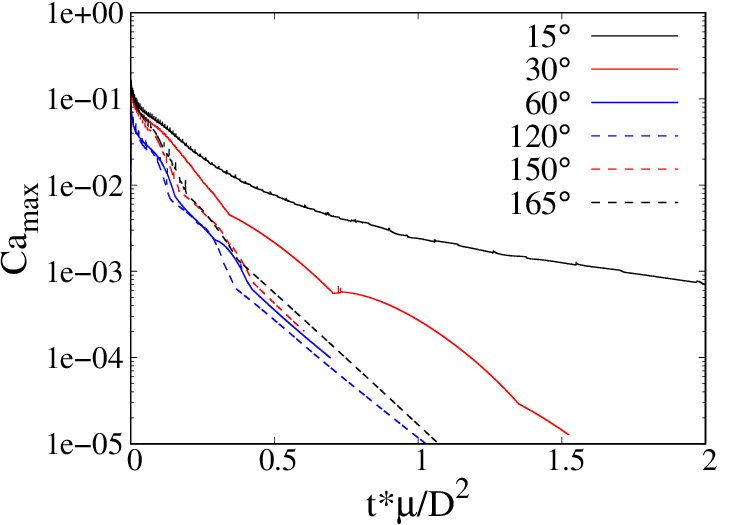}
        \caption{}
        \label{fig:sub3-2-2}
    \end{subfigure}
    \caption{Sessile droplet spreading on an inclined flat solid wall. (a) Equilibrium droplet radius as a function of contact angle. (b) Time evolution of the maximum velocity in the computational domain, defined as $\mathrm{Ca}_{\max} = \mu |u_{\max}|/\sigma$.}
    \label{fig3-2}
\end{figure}

The equilibrium droplet shapes at different contact angles are shown in Fig.~\ref{fig3-1}(b). Between $15^{\circ}$ and $165^{\circ}$, the simulated interface profiles closely match the theoretical circular-cap shapes. The equilibrium radius $R$, obtained by averaging over the entire interface, is plotted as a function of the contact angle in Fig.~\ref{fig3-2}(a), where it coincides with the analytical solution given by Eq.~\eqref{Rthe1}. Moreover, Fig.~\ref{fig3-2}(b) shows the temporal evolution of the maximum velocity in the computational domain, which decays continuously with time. This decay is slower at small contact angles, especially at $15^{\circ}$. As illustrated in Fig.~\ref{fig3-1}(b), the spreading radius increases as the contact angle decreases, implying that more time is required for the droplet to reach equilibrium. As in Sec.~\ref{subsec:stds}, the equilibrium state is identified when the maximum change of the volume fraction between two successive time steps in any cell falls below $10^{-6}$.

\subsubsection{Circular solid surface}

\begin{figure}[t]
    \centering
    \begin{subfigure}[b]{0.33\textwidth}
        \centering
        \includegraphics[width=\textwidth]{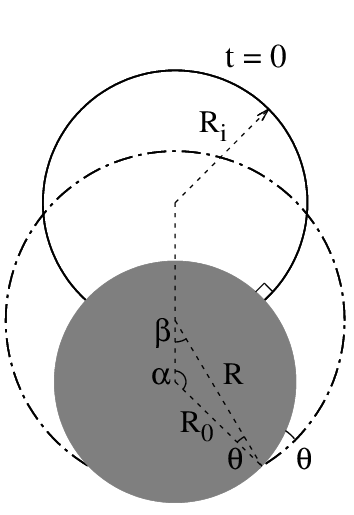}
        \caption{}
        \label{fig:sub3-3-1}
    \end{subfigure}
    \hfill
    \begin{subfigure}[b]{0.52\textwidth}
        \centering
        \includegraphics[width=\textwidth]{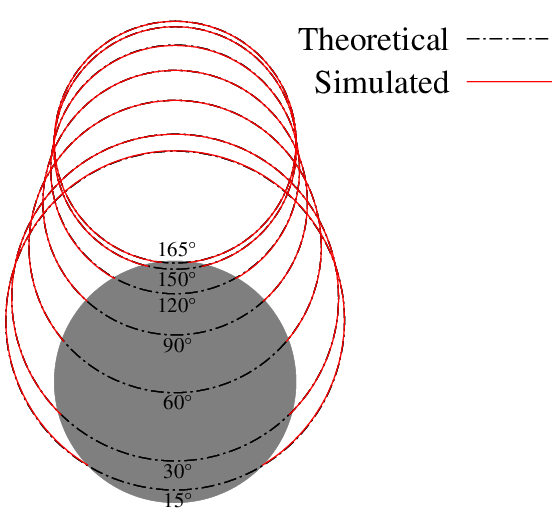}
        \caption{}
        \label{fig:sub3-3-2}
    \end{subfigure}
    \caption{Sessile droplet spreading on a circular cylinder. (a) Schematic diagram. $R_i$ and $R$ denote the initial and equilibrium droplet radii, respectively, while $R_0$ is the cylinder radius. (b) Equilibrium droplet shapes at contact angles of $\theta=15^{\circ}$, $30^{\circ}$, $60^{\circ}$, $90^{\circ}$, $120^{\circ}$, $150^{\circ}$, and $165^{\circ}$. The black dash-dotted lines represent the theoretical circular-cap profiles.}
    \label{fig3-3}
\end{figure}

We next consider a similar droplet spreading problem, but on a stationary circular cylinder. As shown in Fig.~\ref{fig3-3}(a), the cylinder has a radius $R_0 = 0.140\,\text{m}$. A droplet with volume $\pi R_0^2$ is initially placed on the cylinder in the form of a circular cap of radius $R_i = 0.153\,\text{m}$, intersecting the cylinder at $90^{\circ}$. The distance between the initial droplet center and the cylinder center is $d = 0.208\,\text{m}$. A prescribed contact angle $\theta \neq 90^{\circ}$ is then imposed, causing the droplet to spread or retract until reaching a new equilibrium state. Gravity is neglected, so the theoretical equilibrium shape corresponds to a circular cap of radius $R$, consistent with the imposed contact angle. Based on a geometric analysis, the equilibrium radius $R$ can be determined by solving the following system of equations:
\begin{equation}
\alpha + \beta + \theta = 180^{\circ},
\end{equation}
\begin{equation}
R\sin\beta = R_0\sin\alpha,
\end{equation}
\begin{equation}
R^2(\pi-\beta+\sin\beta\cos\beta) + R_0^2(\sin\alpha\cos\alpha-\alpha-\pi) = 0,
\end{equation}
where the definitions of $\alpha$ and $\beta$ are illustrated in Fig.~\ref{fig3-3}(a). The physical parameters of the two fluids are identical to those in the previous subsection \ref{subsubsec-3.3.1}, and the grid resolution is set to 18 cells per cylinder radius $R_0$.

\begin{figure}[t]
    \centering
    \begin{subfigure}[b]{0.49\textwidth}
        \centering
        \includegraphics[width=\textwidth]{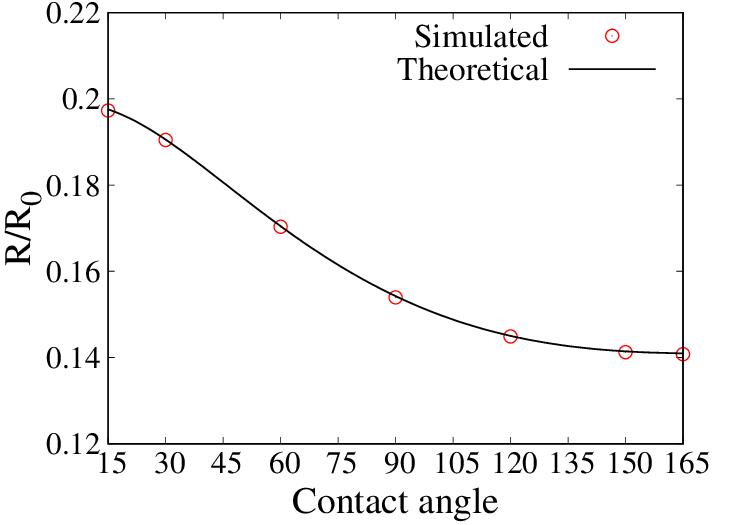}
        \caption{}
        \label{fig:sub3-4-1}
    \end{subfigure}
    \hfill
    \begin{subfigure}[b]{0.49\textwidth}
        \centering
        \includegraphics[width=\textwidth]{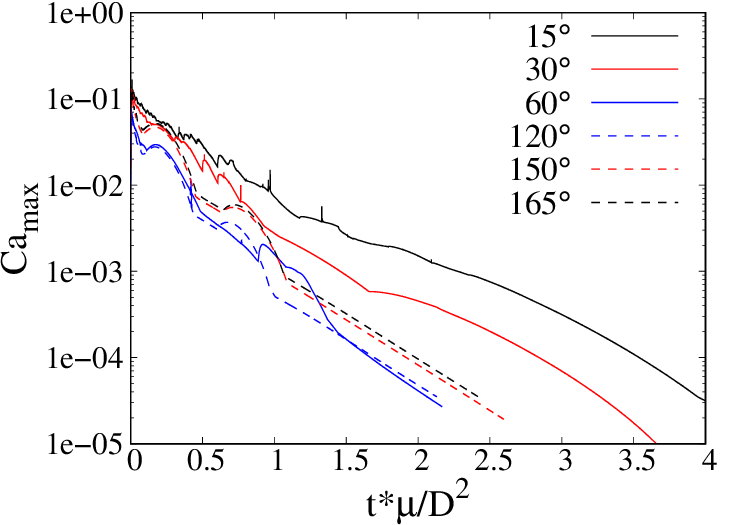}
        \caption{}
        \label{fig:sub3-4-2}
    \end{subfigure}
    \caption{Sessile droplet spreading on a circular cylinder. (a) Variation of the equilibrium droplet radius with contact angle. (b) Time evolution of the maximum velocity in the computational domain, defined as $\mathrm{Ca}_{\max} = \mu |u_{\max}|/\sigma$.}
    \label{fig3-4}
\end{figure}

The equilibrium droplet shapes for contact angles ranging from $15^{\circ}$ to $165^{\circ}$ are shown in Fig.~\ref{fig3-3}(b). The simulated profiles exhibit close agreement with the theoretical circular-cap profiles. The corresponding equilibrium radii, extracted from the simulations, are plotted as a function of the contact angle in Fig.~\ref{fig3-4}(a), where they agree well with the analytical solution. In addition, Fig.~\ref{fig3-4}(b) illustrates that the maximum velocity in the computational domain decays continuously as the droplet relaxes to equilibrium.

For the two test cases examined in this subsection, the equilibrium droplet shapes obtained here show improved accuracy compared with most previous results \cite{patel2017coupled,o2018volume,tavares2024coupled,chen2025volume}, particularly at extreme contact angles. Comparable accuracy was reported only by Huang et al. \cite{huang20252d}, whose contact angle enforcement, however, relies on a more complex parabola fitting scheme. Moreover, to the best of our knowledge, no prior studies have analyzed the velocity evolution in the computational domain, where spurious currents are prone to develop in the vicinity of the contact line due to curvature errors closely related to contact angle treatment. This issue becomes more pronounced for complex solid geometries represented within Cartesian grids using IBM or EBM methods. The present results demonstrate that the proposed contact-angle model effectively suppresses such spurious currents.

\subsection{Droplet penetration into porous media}

\begin{figure}[t]%% placement specifier
	\centering%% For centre alignment of image.
	\includegraphics[width=0.4\textwidth]{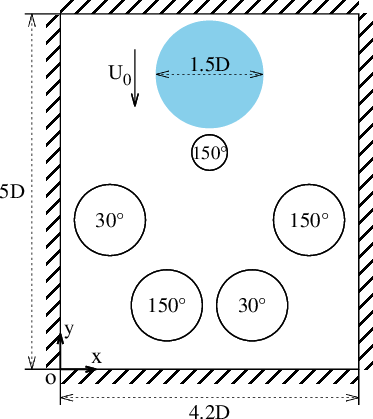}
	\caption{Schematic of droplet penetration into porous media. The four bottom cylinders have a diameter of $D$, and the top cylinder has a radius of $0.5D$. The wettability of each cylinder surface is labeled inside, while the surrounding wall has a contact angle of $90^{\circ}$.}\label{fig-a1}
\end{figure}

In this subsection, we consider a two-dimensional droplet penetrating into a porous medium. This benchmark case has been investigated in several previous studies \cite{liu2015484,o2018volume,huang20252d}. As shown in Fig.~\ref{fig-a1}, a droplet of diameter $1.5D$ is initially positioned at $(2.1D,\,4.05D)$ and falls with an initial velocity $U_0$ into a rectangular domain of size $4.2D\times5D$. Within the domain, five circular cylinders are embedded: one small cylinder of diameter $0.5D$ centered at $(2.1D,\,3.05D)$ with a contact angle of $150^{\circ}$, and four large cylinders of diameter $D$ located at $(0.7D,\,2.1D)$, $(3.5D,\,2.1D)$, $(1.5D,\,0.9D)$, and $(2.7D,\,0.9D)$. The large cylinders have two distinct contact angles, $30^{\circ}$ and $150^{\circ}$, as labelled in Fig.~\ref{fig-a1}, while the surrounding box walls have a contact angle of $90^{\circ}$. Gravity acts downward with acceleration $g$. The physical parameters are set as follows: $D = 1\,\mathrm{mm}$, $\rho_l = 1000\,\mathrm{kg/m^3}$, $\rho_g = 1\,\mathrm{kg/m^3}$, $\mu_l = 1\times10^{-3}\,\mathrm{kg/(m\cdot s)}$, $\mu_g = 2.5\times10^{-5}\,\mathrm{kg/(m\cdot s)}$, and $\sigma = 2.5\times10^{-3}\,\mathrm{N/m}$. Subscripts $l$ and $g$ denote the liquid and gas phases, respectively. The corresponding dimensionless numbers are $\mathrm{Re} = \rho_lU_0D/\mu_l= 50$, $\mathrm{We} = \rho_lU_0^2D/\sigma = 1$, and $\mathrm{Fr} = U_0/\sqrt{gD} = 0.5$. The grid resolution is 85 cells per diameter $D$.

\begin{figure}[tbp]
	\centering
	\begin{tabular}{cccc}
		\includegraphics[width=0.23\textwidth]{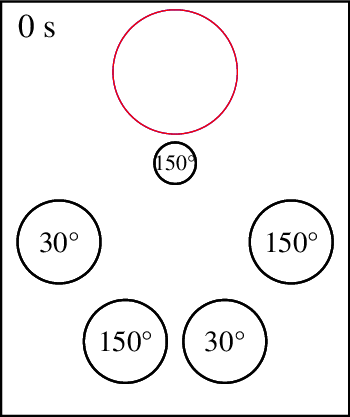} &
		\includegraphics[width=0.23\textwidth]{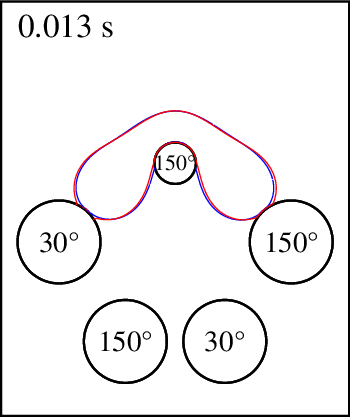} &
		\includegraphics[width=0.23\textwidth]{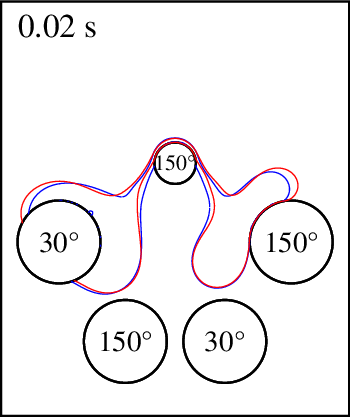} &
		\includegraphics[width=0.23\textwidth]{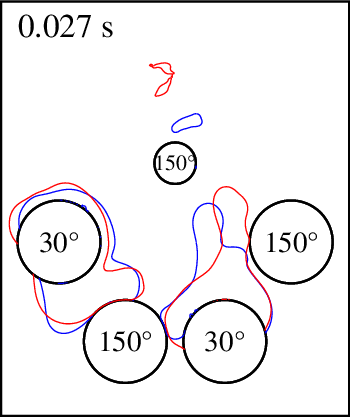} \\
		\includegraphics[width=0.23\textwidth]{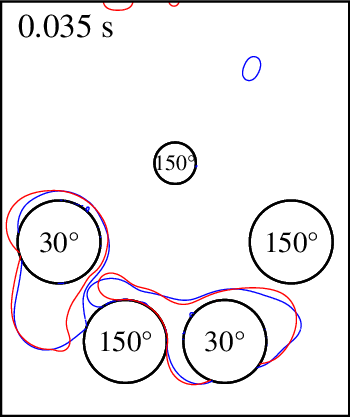} &
		\includegraphics[width=0.23\textwidth]{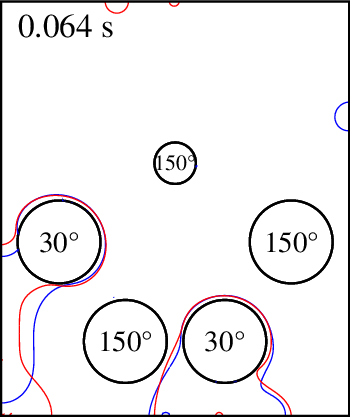} &
		\includegraphics[width=0.23\textwidth]{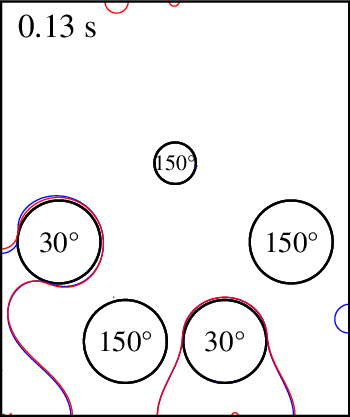} &
		\includegraphics[width=0.23\textwidth]{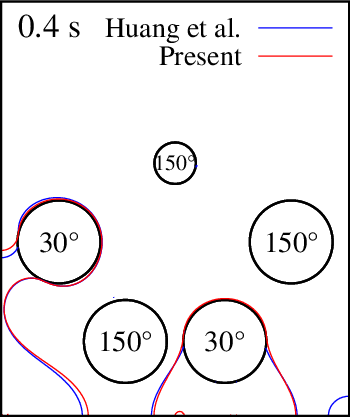} \\
	\end{tabular}
	\caption{Interface profiles at selected time instants. The results of Huang et al.\cite{huang20252d} are also included for comparison.}
	\label{fig-a2}
\end{figure}

The temporal evolution of the interface profiles is shown in Fig.~\ref{fig-a2}. After impacting the small central cylinder, the droplet does not wet its surface but splits into two parts, as seen before $t = 0.02~\mathrm{s}$. The left portion wets the upper hydrophilic cylinder with a contact angle of $30^{\circ}$, whereas the right portion does not adhere to the upper hydrophobic cylinder with a contact angle of $150^{\circ}$. The thin liquid film formed above the small cylinder eventually ruptures and produces a secondary droplet, as observed at $t = 0.027~\mathrm{s}$. Meanwhile, the liquid reaches the two lower cylinders: the hydrophilic cylinder is wetted, while the hydrophobic one remains unwetted. Finally, the liquid becomes trapped between the hydrophilic cylinder and the surrounding walls, forming a stable equilibrium configuration.

The simulated results are also compared with those of Huang et al. \cite{huang20252d}, as shown in Fig.~\ref{fig-a2}. The two solutions exhibit close agreement at the final equilibrium state (after $t = 0.13~\mathrm{s}$), whereas minor discrepancies appear at intermediate times. This may stem from the difference in the procedure used to identify the contact-line cell, as discussed in Sec.~\ref{sec:ecg}.

\subsection{Two-phase flows in microchannels}

In the final subsection, we examine two-phase interfacial flows in microchannels, where surface-tension effects are typically significant. Two channel geometries are considered: a straight channel and a sinusoidal channel. The objective here is to assess the performance of the proposed contact-angle enforcement on embedded solid boundaries and to identify the remaining limitations of the present approach.

\subsubsection{Poiseuille flow in a straight microchannel}
\label{3.5.1}

\begin{figure}[t]%% placement specifier
	\centering%% For centre alignment of image.
	\includegraphics[width=0.4\textwidth]{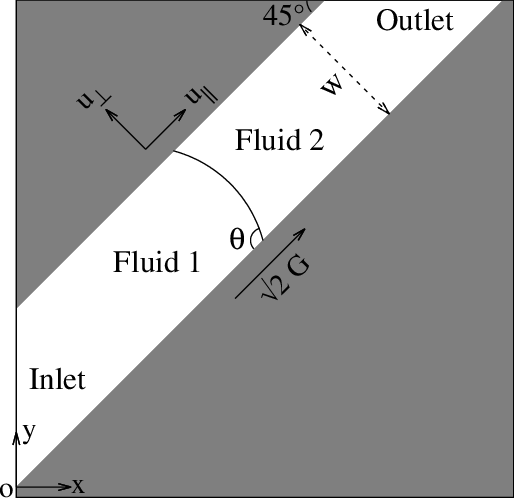}
	\caption{Schematic of two-phase Poiseuille flow in a straight microchannel.}\label{fig4-1}
\end{figure}

As illustrated in Fig.~\ref{fig4-1}, we consider a Poiseuille flow in a straight microchannel, which is embedded within the computational domain and oriented at $45^{\circ}$ with respect to the grid lines. The computational domain is a square of side length $L_0 = 0.4\,\text{mm}$, with the channel inlet located on the left boundary and the outlet on the top boundary. The channel width is $w = 0.102\,\text{mm}$. A two-phase flow is initialized inside the channel, with Fluid~1 displacing Fluid~2. Both fluids are assumed to have identical density and viscosity, $\rho = 1066\,\text{kg}/\text{m}^3$ and $\mu = 0.79\times10^{-3}\,\text{kg}/(\text{m}\cdot\text{s})$, respectively, while the surface-tension coefficient is $\sigma = 0.0657\,\text{N}/\text{m}$. The corresponding dimensionless Laplace number is $\mathrm{La} = {\rho \sigma w}/\mu^2 = 1.14\times10^{4}$, which is relatively 
large so as to magnify the influence of surface tension forces. A constant contact angle of $\theta = 120^{\circ}$ is imposed at the contact line. At both the inlet and outlet, a parabolic velocity profile is prescribed as
\begin{equation}
u_x = u_y = \frac{\rho G}{4\mu}y(\sqrt{2}w-y),
\end{equation}
where $G = 3.75\,\text{m}/\text{s}^2$ denotes the body-force acceleration applied to both fluids in the $x$- and $y$-directions. A no-slip velocity condition is imposed on all embedded solid walls. In this case, the capillary number is defined as $\mathrm{Ca} = \mu u_{\max}/\sigma = 1.12\times10^{-4}$, where $u_{\max} = \sqrt{2}\rho G w^2/(8\mu)$ denotes the maximum velocity magnitude at the channel inlet. The Reynolds number is defined as $\mathrm{Re} = \mathrm{Ca}\cdot\mathrm{La} = \rho u_{\max} w/\mu = 1.27$. The grid resolution is set to 16 cells per channel width.

\begin{figure}[tbp]
    \centering
    \begin{subfigure}[b]{0.49\textwidth}
        \centering
        \includegraphics[width=\textwidth]{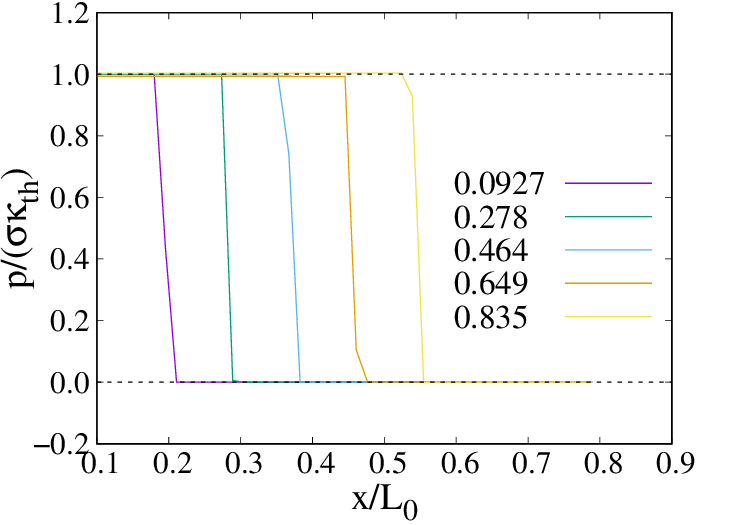}
        \caption{}
        \label{fig:sub4-2-1}
    \end{subfigure}
    \hfill
    \begin{subfigure}[b]{0.49\textwidth}
        \centering
        \includegraphics[width=\textwidth]{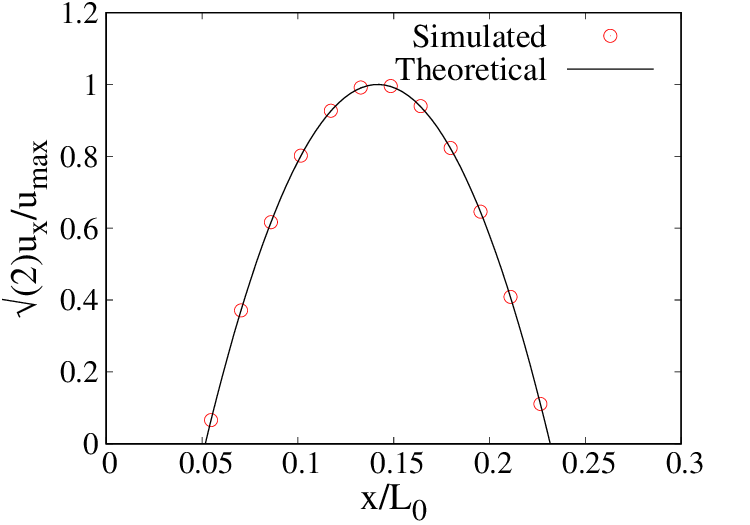}
        \caption{}
        \label{fig:sub4-2-2}
    \end{subfigure}
    \caption{Two-phase Poiseuille flow in a straight microchannel at $\mathrm{La} = 1.14\times10^{4}$ and $\mathrm{Ca} = 1.12\times10^{-4}$. (a) Normalized pressure distribution along the channel centerline at five time instants: $t^* = t u_{\max}/L_0 = 0.0927$, $0.278$, $0.464$, $0.649$ and $0.835$. (b) Normalized velocity distribution $u_x/(u_{\max}/\sqrt{2})$ at a cross-section located far from the fluid–fluid interface at $t^* = 0.742$.}
    \label{fig4-2}
\end{figure}

The simulation results are shown in Fig.~\ref{fig4-2}. Fig.~\ref{fig4-2}(a) plots the pressure distribution along the channel centerline ($y = x + w/\sqrt{2}$) at different time instants. A distinct pressure drop can be observed at the fluid–fluid interface, which propagates at a constant speed. This pressure discontinuity arises from the surface-tension force, expressed as $\sigma \kappa_{th}$, where the theoretical interface curvature is $\kappa_{th} = 1/w$. As shown in Fig.~\ref{fig4-2}(a), the simulated pressure drop agrees well with the theoretical prediction.

\begin{figure}[t]
    \centering
    \begin{subfigure}[b]{0.4\textwidth}
        \centering
        \includegraphics[width=\textwidth]{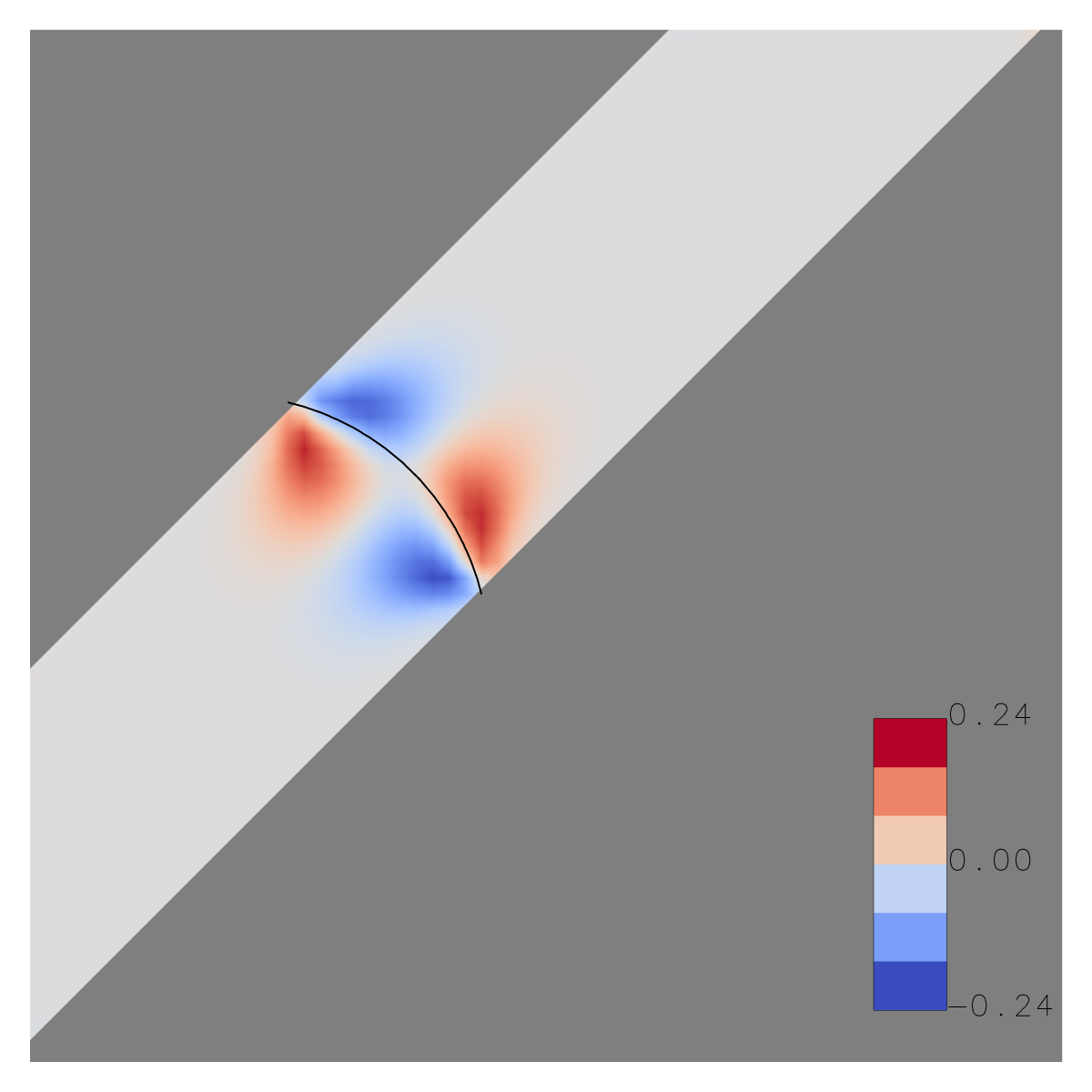}
        \caption{}
        \label{fig:sub4-1-1}
    \end{subfigure}
    \hspace{1cm}
    \begin{subfigure}[b]{0.4\textwidth}
        \centering
        \includegraphics[width=\textwidth]{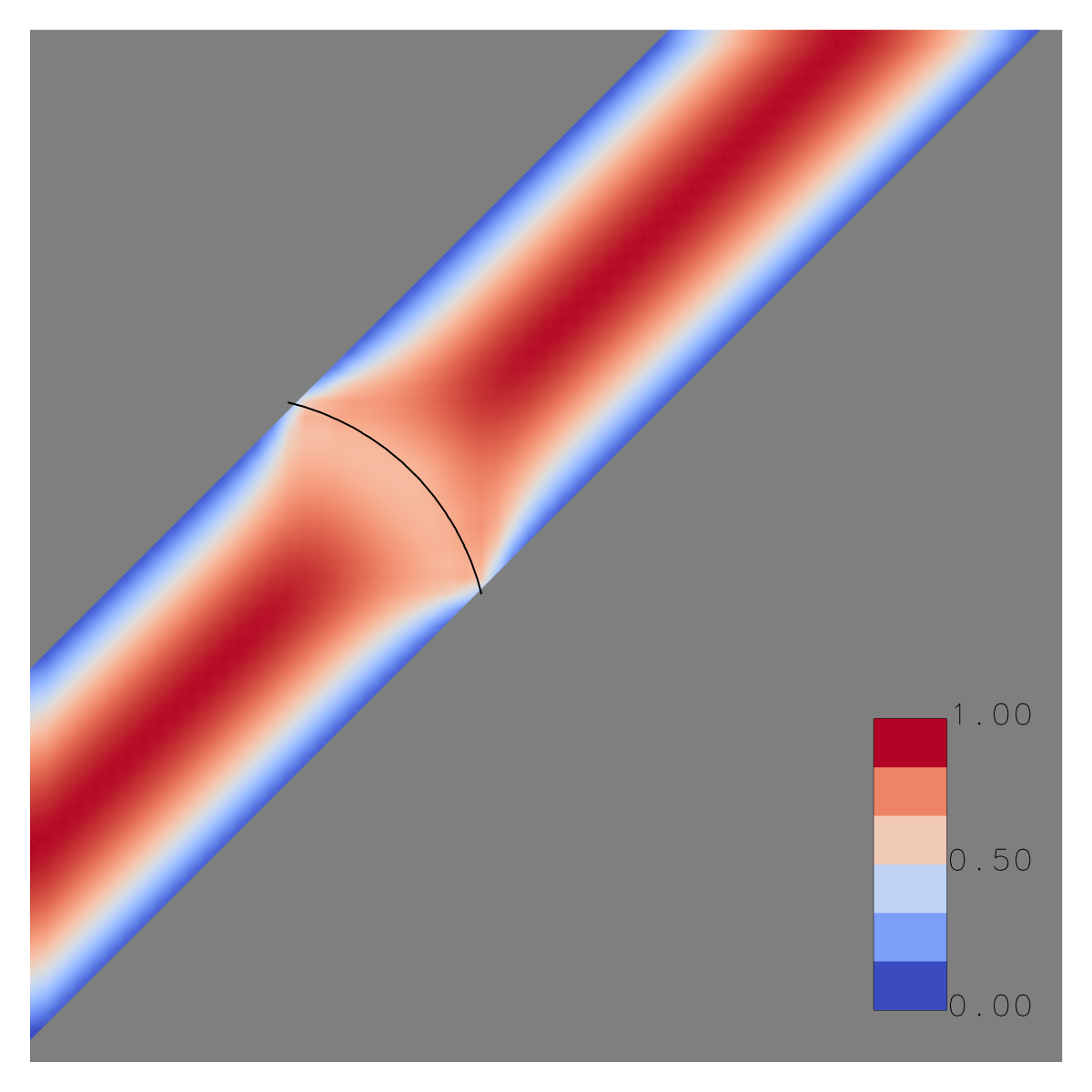}
        \caption{}
        \label{fig:sub4-1-2}
    \end{subfigure}
    \caption{Contours of the normalized velocity components at $\mathrm{La} = 1.14\times10^{4}$ and $\mathrm{Ca} = 1.12\times10^{-4}$: (a) $u_{\perp}/u_{\max}$ and (b) $u_{\parallel}/u_{\max}$. Here, $u_{\perp}$ and $u_{\parallel}$ denote the velocity components normal and parallel to the wall. An animation showing the temporal evolution of $u_{\parallel}/u_{\max}$ is available in the supplementary material \cite{data2025}.}
    \label{figa-3}
\end{figure}

The contours of the normalized velocity components, $u_{\parallel}/u_{\max}$ and $u_{\perp}/u_{\max}$, are shown in Fig.~\ref{figa-3}. Here, $u_{\parallel}$ and $u_{\perp}$ denote the velocity components parallel and perpendicular to the solid wall, respectively, as illustrated in Fig.~\ref{fig4-1}. Away from the fluid–fluid interface, the velocity distribution across the channel cross-section agrees well with the theoretical parabolic profile, as shown in Fig.~\ref{fig4-2}(b). In the vicinity of the interface, deviations from the parabolic velocity profile are observed, which are physically necessary for maintaining a uniform translation of the interface within the channel.

\begin{figure}[tbp]
    \centering
    \begin{subfigure}[b]{0.49\textwidth}
        \centering
        \includegraphics[width=\textwidth]{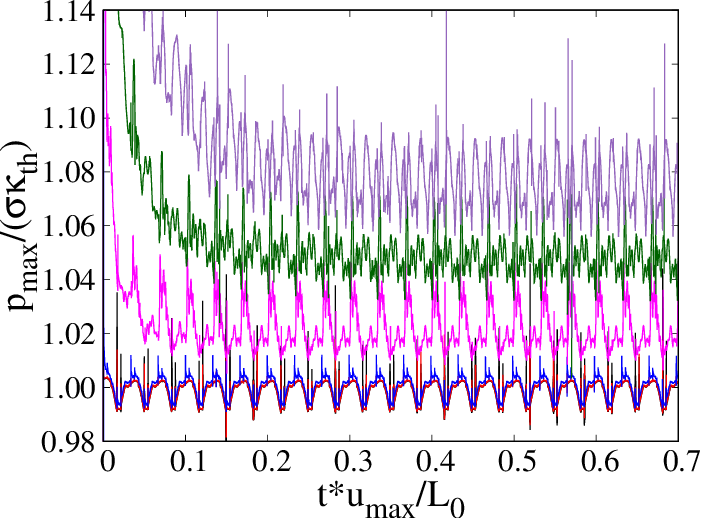}
        \caption{}
    \end{subfigure}
    \hfill
    \begin{subfigure}[b]{0.49\textwidth}
        \centering
        \includegraphics[width=\textwidth]{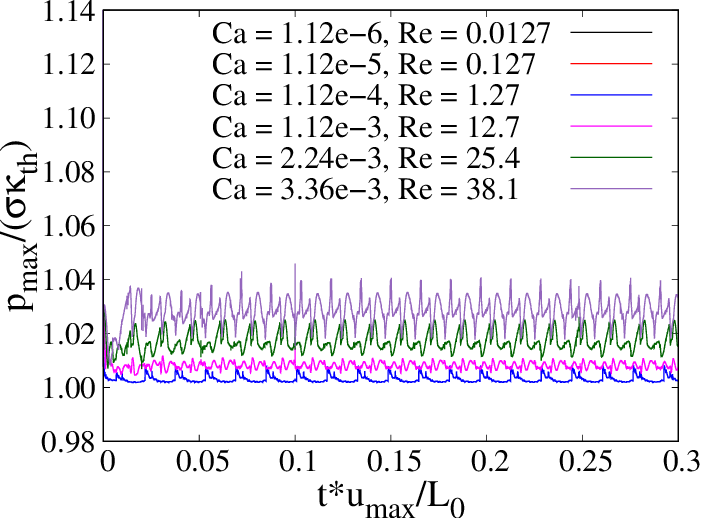}
        \caption{}
    \end{subfigure}
    \caption{Temporal evolution of the normalized maximum pressure for different capillary numbers at $\mathrm{La}=1.14\times10^{4}$. (a) Straight microchannel with inclined walls. (b) Reference case: an entire droplet advected by a uniform flow in the absence of solid walls.}
    \label{fig4-3}
\end{figure}

The temporal evolution of the maximum pressure in the channel for various capillary numbers is shown in Fig.~\ref{fig4-3}(a). The capillary number $\mathrm{Ca}$ is controlled by varying the body-force acceleration $G$. When $\mathrm{Ca} < 1.12\times10^{-4}$, the results almost collapse onto a single curve, indicating that surface tension dominates and the interface remains close to a circular arc of radius $w$. A slight periodic oscillation is also observed, with an amplitude below 2\% of the mean pressure. This oscillation can be attributed to the curvature evaluation within the contact-line cell. Recall that the curvature in this cell is computed using Eqs.~\eqref{hxcl}–\eqref{hxxcl}, which explicitly depend on the instantaneous contact-line position. Consequently, as the contact line moves within a cut cell, the evaluated curvature varies continuously, and this variation pattern repeats each time the contact line crosses into an adjacent cut cell.

With increasing $\mathrm{Ca}$, the mean maximum pressure increases as expected, due to the reduced influence of surface tension and the enhanced viscous effects at higher $\mathrm{Ca}$. The interface becomes more susceptible to flow-induced deformations, thereby deviating progressively from the theoretical circular-arc shape. Moreover, when $\mathrm{Ca} > 2.24\times10^{-3}$, the pressure oscillations become more irregular, as shown in Fig.~\ref{fig4-3}(a), owing to the higher Reynolds number and the associated inertial effects.

\begin{figure}[tbp]
	\centering
	\begin{subfigure}[b]{0.49\textwidth}
		\centering
		\includegraphics[width=\textwidth]{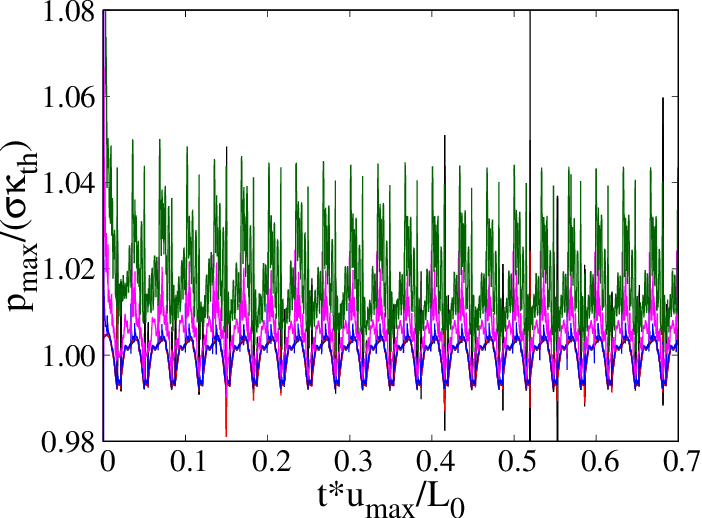}
		\caption{}
	\end{subfigure}
	\hfill
	\begin{subfigure}[b]{0.49\textwidth}
		\centering
		\includegraphics[width=\textwidth]{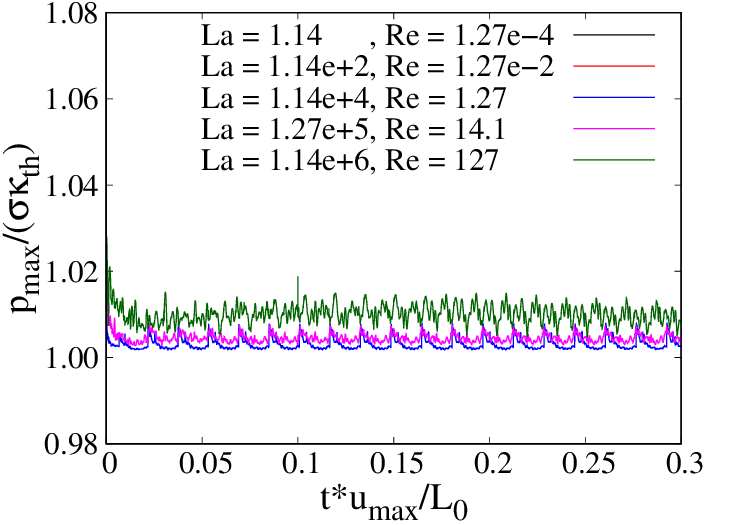}
		\caption{}
	\end{subfigure}
	\caption{Temporal evolution of the normalized maximum pressure for different Laplace numbers at $\mathrm{Ca}=1.12\times10^{-4}$. (a) Straight microchannel with inclined walls. (b) Reference case: an entire droplet advected by a uniform flow in the absence of solid walls.}
	\label{fig4-4}
\end{figure}

To assess the influence of contact-angle enforcement on the observed pressure oscillations, Fig.~\ref{fig4-3}(b) presents the temporal evolution of the maximum pressure for a series of reference cases computed using the same parameters as in Fig.~\ref{fig4-3}(a). In the reference cases, an entire droplet is advected by a uniform flow throughout the computational domain shown in Fig.~\ref{fig4-1}(a), without any solid boundaries. This setup was originally proposed by Popinet \cite{popinet2009accurate} to investigate the spurious currents arising from the discretization of surface tension combined with interface advection. The results for $\mathrm{Ca} < 1.12\times10^{-5}$ are omitted, since under such small $\mathrm{Ca}$ the magnitude of spurious flows becomes comparable to or even exceeds that of the imposed uniform flow, thereby preventing droplet translation. For all reported $\mathrm{Ca}$, the pressure oscillations are significantly smaller than those in Fig.~\ref{fig4-3}(a), indicating that contact-angle enforcement is the primary source of the oscillations observed in the embedded-channel configuration. Moreover, when $\mathrm{Ca}>2.24\times10^{-3}$, inertial effects become evident, consistent with the behavior seen in Fig.~\ref{fig4-3}(a).

Fig.~\ref{fig4-4}(a) compares the temporal evolution of the maximum pressure oscillation for different Laplace numbers, which are varied by adjusting the fluid viscosity. When $\mathrm{La} < 1.14\times10^{4}$, the pressure-oscillation curves nearly overlap. Although viscous effects become more pronounced with decreasing $\mathrm{La}$, the flow velocity in the channel decreases simultaneously due to the increasing fluid viscosity. As a result, no significant interface deformation induced by the flow is observed. In contrast, when $\mathrm{La} > 1.27\times10^{5}$, the pressure oscillations become increasingly irregular due to the rapidly growing Reynolds number, which amplifies inertial effects. Fig.~\ref{fig4-4}(b) presents the corresponding results for the reference cases, showing that the pressure fluctuations induced by pure interface advection are much smaller than those associated with the contact-angle enforcement. Extended pressure‐oscillation statistics over a wider range of parameters are provided in Table~\ref{tab:Ca_Oh} of Appendix~\ref{app1}. The corresponding velocity fluctuations in this embedded microchannel flow are further examined in Appendix~\ref{app2}.

\subsubsection{Pressure-driven flow in a sinusoidal microchannel}
\label{3.5.2}

\begin{figure}[t]%% placement specifier
	\centering%% For centre alignment of image.
	\includegraphics[width=0.7\textwidth]{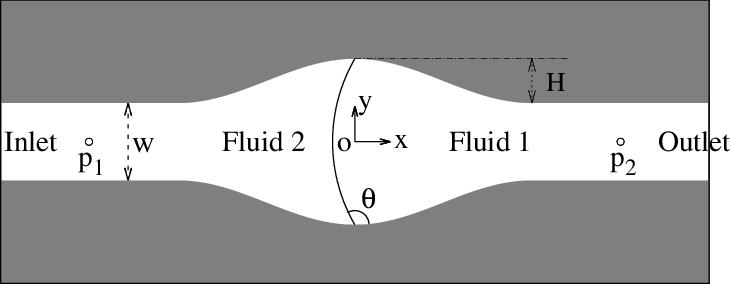}
	\caption{Schematic of pressure-driven two-phase flow in a sinusoidal microchannel.}\label{fig5-1}
\end{figure}

We consider a pressure-driven two-phase flow in a sinusoidal microchannel, as illustrated in Fig.~\ref{fig5-1}. The channel has a total length of $L_0 =0.4\,\text{mm}$. The top and bottom walls of the channel are symmetric about $y=0$, and the shape of the top wall is defined as
\begin{equation}
y =
\begin{cases}
\dfrac{w}{2}, & \text{if } x < -\dfrac{L_0}{4} \text{ or } x > \dfrac{L_0}{4}, \\[6pt]
\dfrac{w}{2} + \dfrac{H}{2}\left(1 + \cos\left(\dfrac{4\pi x}{L_0}\right)\right), & \text{if } -\dfrac{L_0}{4} \leq x \leq \dfrac{L_0}{4}.
\end{cases}
\end{equation}
where $w = 0.0375\,\text{mm}$ is the width of the straight section and $H = 0.025\,\text{mm}$ is the amplitude of the sinusoidal section, as shown in Fig.~\ref{fig5-1}. A two-phase flow is initialized in the straight inlet section, with Fluid~2 displacing Fluid~1. The physical properties of both fluids are identical to those used in Sec.~\ref{3.5.1}, corresponding to an Laplace number of $\mathrm{La} = {\rho \sigma w}/\mu^2 = 4.21\times10^{3}$. A constant contact angle of $\theta = 120^{\circ}$ is imposed at the contact line. At the inlet, a uniform velocity $U_0=1\,\text{cm/s}$ is prescribed, while a free-outflow boundary condition is applied at the outlet. No-slip conditions are imposed on all embedded solid boundaries. The capillary number in this case is defined as $\mathrm{Ca} = \mu U_0/\sigma = 1.20\times10^{-4}$. The grid resolution is set to 12 cells per channel width $w$.

\begin{figure}[tbp]
	\centering
	\begin{subfigure}[b]{0.49\textwidth}
		\centering
		\includegraphics[width=\textwidth]{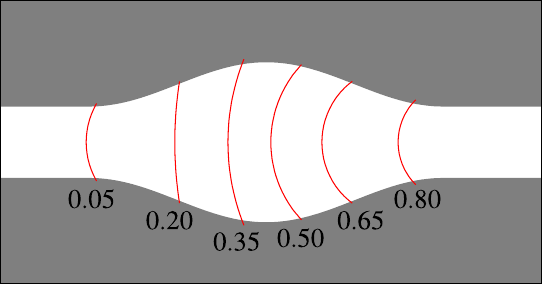}
		\caption{}
	\end{subfigure}
	\hfill
	\begin{subfigure}[b]{0.49\textwidth}
		\centering
		\includegraphics[width=\textwidth]{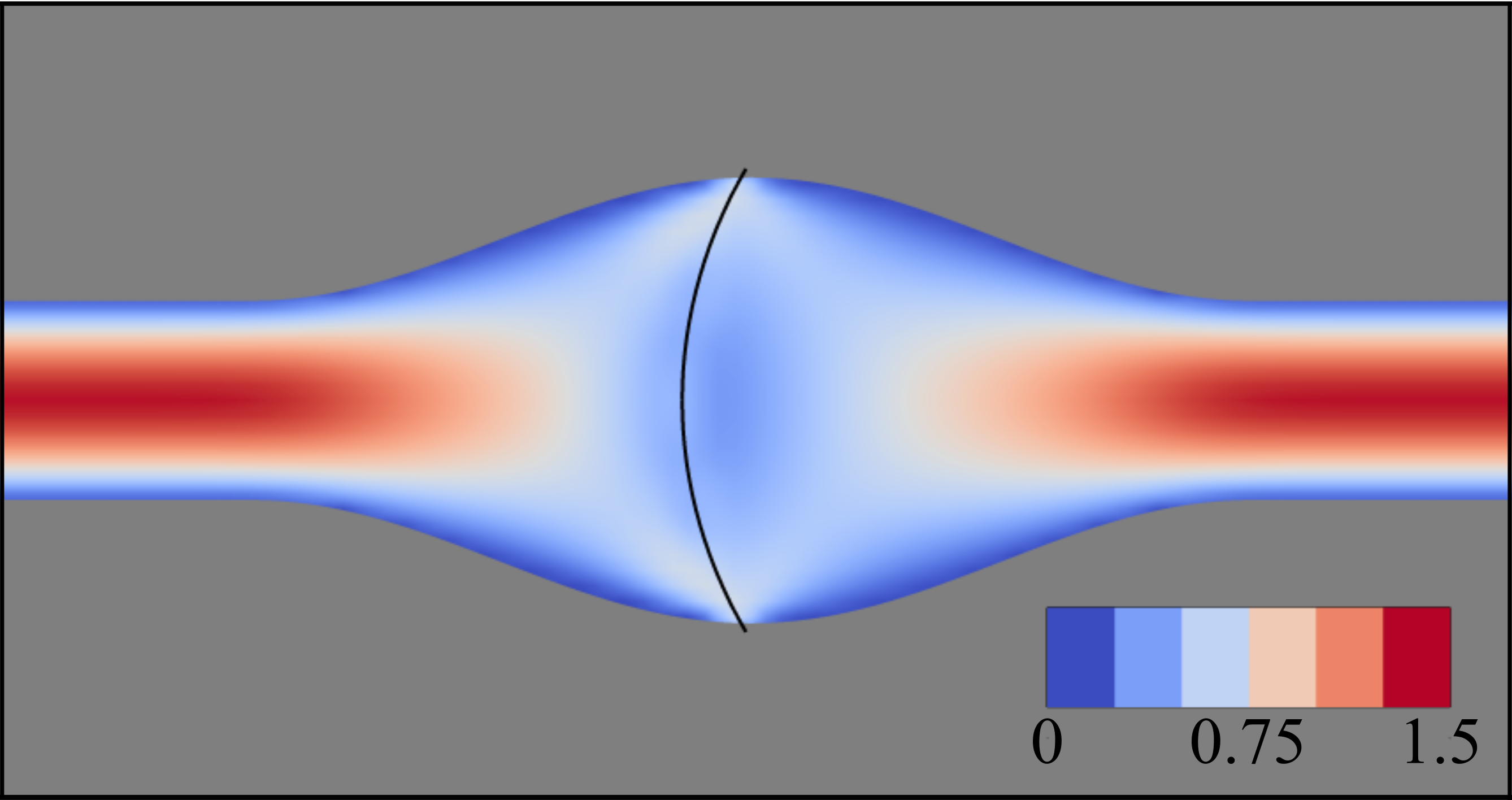}
		\caption{}
	\end{subfigure}
	\caption{Two-phase pressure-driven flow in a sinusoidal microchannel at $\mathrm{La} = 4.21\times10^{3}$ and $\mathrm{Ca} = 1.20\times10^{-4}$. (a) Interface profiles at six time instants, $t^* = t U_{0}/L_0 = 0.05$, $0.20$, $0.35$, $0.50$, $0.65$, and $0.80$. (b) Contours of the normalized streamwise velocity $u_x/U_0$ at $t^* = 0.4$.}
	\label{fig5-2}
\end{figure}

The simulation results are shown in Fig.~\ref{fig5-2}. Fig.~\ref{fig5-2}(a) presents the interface profiles at several time instants. As the interface enters the sinusoidal section, its curvature varies with the local wall orientation owing to the prescribed constant contact-angle condition. The interface curvature is noticeably smaller in the expansion region than in the contraction region. Fig.~\ref{fig5-2}(b) shows the contours of the normalized streamwise velocity, $u_x/U_0$, at $t^*= t U_{0}/L = 0.4$, when the interface reaches the center of the channel. In the straight inlet and outlet sections, the velocity profile across the channel cross-section agrees well with the theoretical parabolic distribution. In contrast, within the sinusoidal section, the velocity field deviates from the parabolic profile due to local variations in cross-sectional area as well as the presence of the interface.

\begin{figure}[t]%% placement specifier
	\centering%% For centre alignment of image.
	\includegraphics[width=0.6\textwidth]{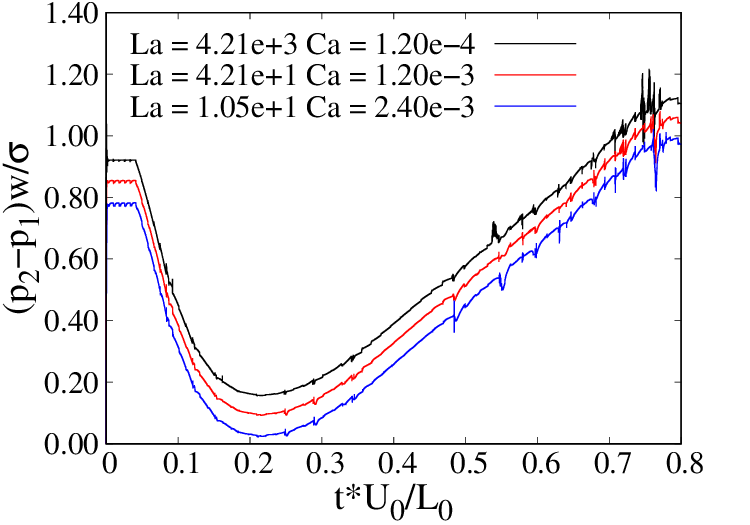}
	\caption{Temporal evolution of the pressure difference $(p_2 - p_1)$ in the sinusoidal microchannel for different parameter sets, with the locations of $p_1$ and $p_2$ indicated in Fig.~\ref{fig5-1}.}\label{fig5-3}
\end{figure}

Moreover, as seen from the interface profiles in Fig.~\ref{fig5-2}(a), the interface motion in the expansion and contraction regions of the sinusoidal section is clearly asymmetric. To further illustrate this difference, Fig.~\ref{fig5-3} shows the temporal evolution of the pressure difference, $p_2 - p_1$, between two reference points located at the centers of the straight inlet and outlet sections, as indicated in Fig.~\ref{fig5-1}. In the expansion region, the interface curvature decreases with time, resulting in smooth contact-line motion. In contrast, in the contraction region, the curvature increases with time, which slows down the interface motion and gives rise to small oscillations. These oscillations are likely associated with fluctuations in the curvature evaluated within the contact-line cell, which appear to be amplified as the interface moves across the contraction region.

Fig.~\ref{fig5-3} also includes two additional results with higher fluid viscosity, corresponding to lower Laplace numbers and higher capillary numbers. The overall variation of the pressure difference, as well as the oscillations observed in the contraction region, remain nearly identical across the three parameter sets. Only the magnitude of $(p_2 - p_1)$ increases with $\mathrm{La}$, reflecting the enhanced influence of surface tension, which makes the interface less susceptible to the flow-induced deformation. The pressure oscillations observed here are consistent with those in Fig.~\ref{fig4-4}, further suggesting that they are associated with the curvature fluctuations within the contact-line cell.

The two test cases considered in this subsection highlight the remaining limitations of the proposed contact angle model when applied to embedded solid boundaries. To the best of our knowledge, no prior studies have conducted similar tests, and this issue seems to be ubiquitous in contact-angle enforcement within geometric VOF formulation that employ sharp representations of solid surfaces, such as the EBM framework. The resulting velocity and pressure fluctuations near the contact line may become non-negligible in flows where surface-tension effects are dominant.

\section{Conclusions}

In this work, we have developed a height-function-based numerical method for enforcing contact angles over the full range on both flat and curved solid surfaces within two-dimensional VOF simulations. The method incorporates the contact-line position into the curvature estimation in contact-line cells, where the interface normal is constrained to satisfy the prescribed contact angle with the solid surface, thereby ensuring smooth motion of the contact line. A key advantage of the proposed method is its straightforward extension from flat to curved solid surfaces represented by the embedded boundary method, while retaining the simplicity of implementation. This extension is made possible by first addressing the challenges of interface reconstruction and geometric advection in cut cells with polygonal fluid regions. The codes developed in this work are freely available in the Basilisk sandbox \cite{sandbox}.

Numerical tests have confirmed the accuracy and robustness of the proposed method. First, an interface advection test around a cylinder validates the reconstruction and advection schemes developed for cut cells. On flat solid surfaces, simulations of surface-tension-driven droplet spreading demonstrate that the method attains higher accuracy than the conventional vertical height-function model when enforcing very small or very large contact angles. Next, contact-angle enforcement on embedded solid boundaries is examined through sessile droplet spreading on inclined flat and circular surfaces, showing that the method can accurately impose arbitrary contact angles without generating significant spurious currents. The simulation of two-dimensional droplet penetration into porous media further verifies the capability of the present method to handle complex solid geometries. Finally, two-phase flows in both straight and sinusoidal microchannels are simulated, revealing the remaining limitations of the present approach when applied to embedded solid boundaries.

Finally, we comment on the possible extension of the proposed 2D contact-angle enforcement strategy to 3D. The key idea---avoiding interface extrapolation from contact line cells into the solid---remains attractive in 3D, as the ghost-cell volume fractions could, in principle, be defined using the information from multiple neighboring contact-line cells simultaneously. However, the extension is not straightforward, particularly for embedded solid boundaries. A major challenge lies in selecting appropriate height-function data within the fluid domain to complete the curvature estimation in contact line cells. In addition, the interface reconstruction in 3D cut cells is inherently more complex than that in 2D.

\section*{Funding}

This project has received funding from the European Research Council (ERC) under the European Union’s Horizon 2020 research and innovation programme (project TRUFLOW, grant agreement number 883849). We are grateful to GENCI for generous allocation on Adastra supercomputers (grant agreement number A0152B14629). This work has been supported by the National Natural Science Foundation of China (Grants No.~12302343 and No.~12372266), and by the Postdoctoral Science Foundation of China (2023M733461).

\section*{Acknowledgment}

The authors gratefully acknowledge Arthur Moncorge, Hamdi Tchelepi, Moataz Abu AlSaud, Martin Blunt, Branko Bijlelic, Mossayeb Shams and Ali Raeni for insightful discussions leading to the test case of contact-line motion in a sinusoidal microchannel in Sec.~\ref{3.5.2}.

\appendix

\renewcommand\thesection{\Alph{section}}

\renewcommand{\thetable}{A.\arabic{table}}
\setcounter{table}{0}

\renewcommand{\thefigure}{A.\arabic{figure}}
\setcounter{figure}{0}

\begin{figure}[H]%% placement specifier
	\centering%% For centre alignment of image.
	\includegraphics[width=0.65\textwidth]{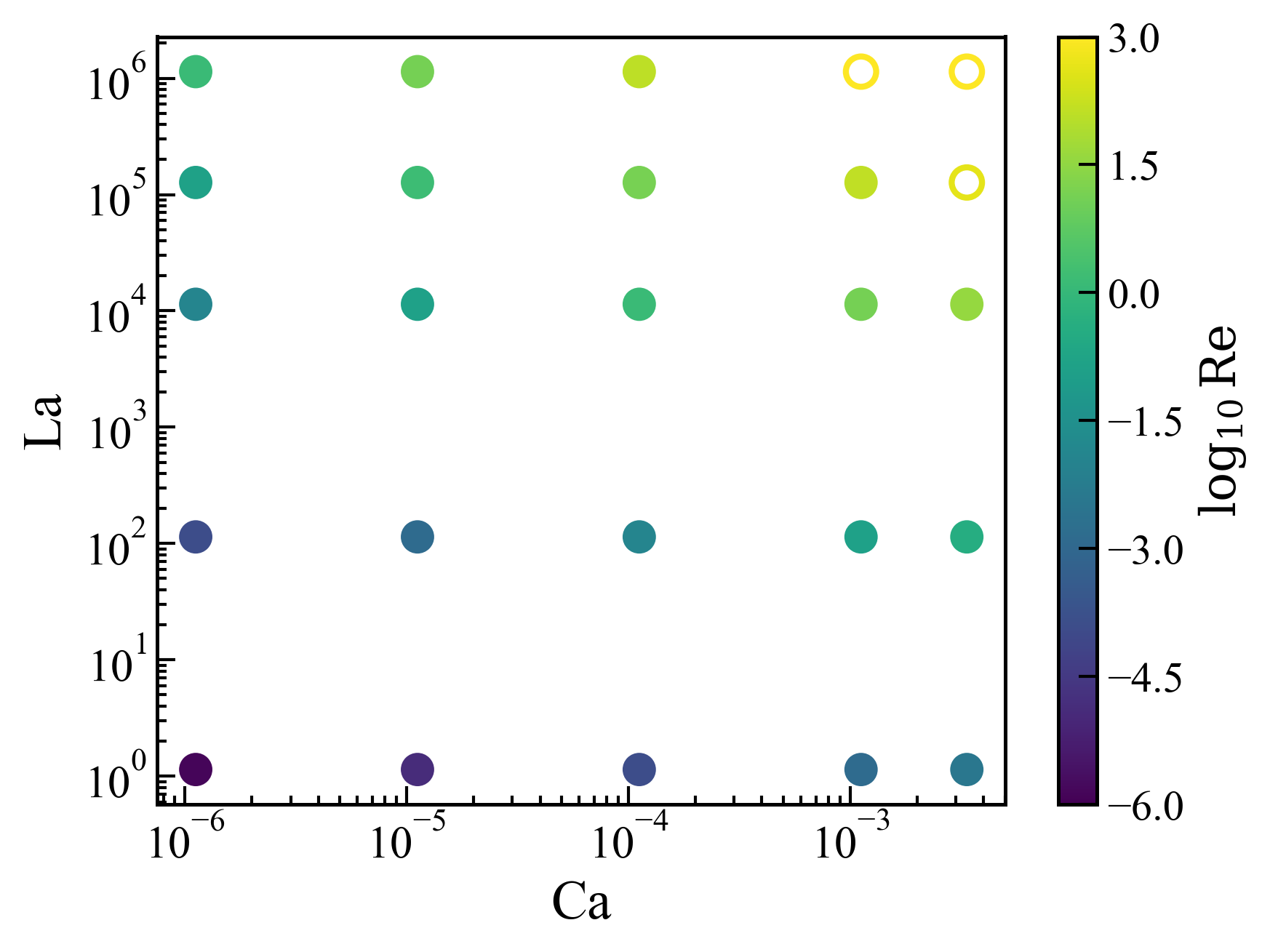}
	\caption{Parameter space of the simulations in Appendix~\ref{app1} and \ref{app2}, showing the combinations of capillary and Laplace numbers, as well as the corresponding Reynolds numbers ($\mathrm{Re}=\mathrm{Ca}\cdot\mathrm{La}$). Hollow points indicate cases that are excluded from the present study.}\label{figa-4}
\end{figure}

\section{Pressure oscillations over a wider parameter space}
\label{app1}

To complement the analysis presented in Sec.~\ref{3.5.1}, the oscillation of the maximum pressure in the straight microchannel (see Fig.~\ref{fig4-1}) is examined over a wider parameter space, as shown in Fig.~\ref{figa-4}. The hollow points in the upper-right region of the figure are excluded from the present study because their Reynolds numbers are excessively large (exceeding 1000), leading to inertia-dominated pressure oscillations overwhelming those associated with contact-angle enforcement. For the remaining cases, the root-mean-square (RMS) value of the maximum pressure fluctuation is computed as:
\begin{equation}\label{a1}
\mathrm{RMS}(p_{\max})
= \sqrt{\frac{1}{N}\sum_{i=1}^{N}
\left(p_{\max,i}-\overline{p}_{\max}\right)^{2}},
\end{equation}
where $p_{\max,i}$ is the instantaneous maximum pressure at time step $i$, and $\overline{p}_{\max}$ denotes its time-averaged value. The results are summarized in Table~\ref{tab:Ca_Oh}. Overall, the pressure fluctuations remain of the same order of magnitude for most parameter combinations, except for those in the lower-right region of the table, where the corresponding Reynolds numbers are relatively high.

\begin{table}[t]
	\centering
	\caption{RMS values of the maximum pressure fluctuations for a Poiseuille-type inflow in an embedded microchannel with moving contact lines.}
	\begin{tabular}{c|ccccc}
		\hline
		\hline
		\multicolumn{1}{c|}{$\mathrm{Ca} \backslash \mathrm{La}$} 
		& $1.14$ 
		& $1.14\times10^{2}$ 
		& $1.14\times10^{4}$ 
		& $1.27\times10^{5}$ 
		& $1.14\times10^{6}$ \\
		\hline
		$1.12\times10^{-6}$ & 0.0128 & 0.00384 & 0.00377 & 0.00404 & 0.00418 \\
		$1.12\times10^{-5}$ & 0.00409 & 0.00385 & 0.00373 & 0.00358 & 0.00344 \\
		$1.12\times10^{-4}$ & 0.00405 & 0.00378 & 0.00383 & 0.00549 & 0.00926 \\
		$1.12\times10^{-3}$ & 0.00380 & 0.00398 & 0.00824 & 0.0121  & ---    \\
		$3.36\times10^{-3}$ & 0.00360 & 0.00501 & 0.00947 & ---     & ---    \\
		\hline
		\hline
	\end{tabular}
	\label{tab:Ca_Oh}
\end{table}

\section{Velocity oscillations in an embedded microchannel}
\label{app2}

\renewcommand{\thetable}{B.\arabic{table}}
\setcounter{table}{0}

\renewcommand{\thefigure}{B.\arabic{figure}}
\setcounter{figure}{0}

\begin{figure}[t]
	\centering
	\begin{subfigure}[b]{0.49\textwidth}
		\centering
		\includegraphics[width=\textwidth]{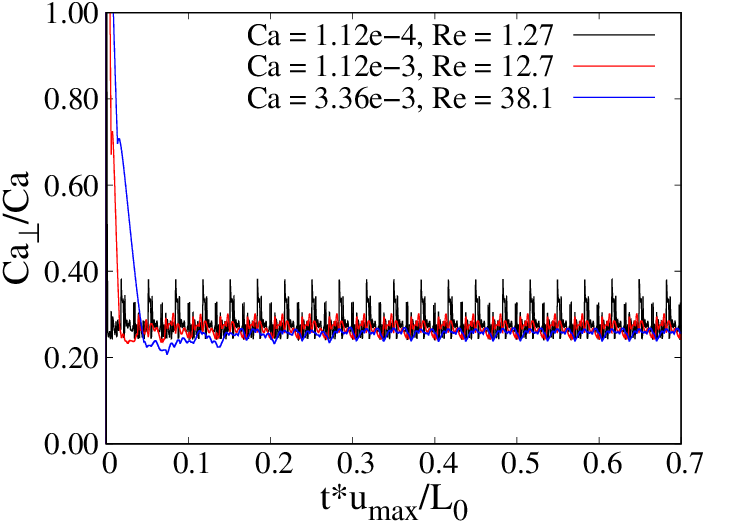}
		\caption{}
	\end{subfigure}
	\hfill
	\begin{subfigure}[b]{0.49\textwidth}
		\centering
		\includegraphics[width=\textwidth]{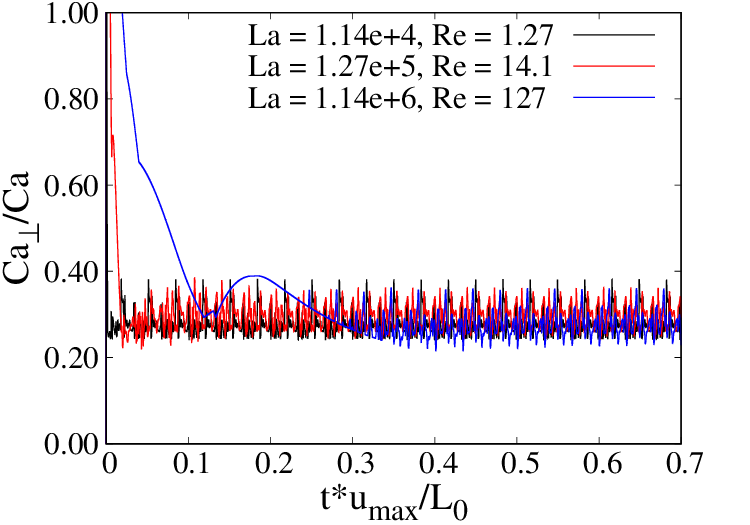}
		\caption{}
	\end{subfigure}
	\caption{Temporal evolution of the ratio $\mathrm{Ca}_{\perp}/\mathrm{Ca}$ for a Poiseuille-type inflow in an embedded microchannel with moving contact lines: (a) different capillary numbers at $\mathrm{La} = 1.14\times10^{4}$ and (b) different Laplace numbers at $\mathrm{Ca}=1.12\times10^{-4}$.}
	\label{figa-5}
\end{figure}

In addition to the pressure oscillations discussed above, the embedded microchannel flow in Sec.~\ref{3.5.1} (see Fig.~\ref{fig4-1}) also exhibits accompanying velocity oscillations, as shown in the supplementary animations \cite{data2025}. To quantify these, the temporal evolution of the ratio $\mathrm{Ca}_{\perp}/\mathrm{Ca}$ is plotted in Fig.~\ref{figa-5}. Here, $\mathrm{Ca}_{\perp}$ and $\mathrm{Ca}$ are calculated as
\begin{equation}
\mathrm{Ca}_{\perp}=\mu |u_{\perp}|_{\max}/\sigma, \,\,\,\,\,\,
\mathrm{Ca}=\mu u_{\max}/\sigma,
\end{equation}
representing the capillary numbers defined using the maximum wall-normal velocity component $|u_{\perp}|_{\max}$, which occurs near the interface as illustrated in Fig.~\ref{figa-3}, and the maximum inlet velocity magnitude $u_{\max}$, respectively. The presence of a nonzero wall-normal velocity near the interface is necessary to maintain its uniform translation along the channel. In both Fig.~\ref{figa-5}(a) and (b), the mean value of $\mathrm{Ca}_{\perp}/\mathrm{Ca}$ remains nearly constant across the tested combinations of $\mathrm{Ca}$ and $\mathrm{La}$, indicating the dominant role of capillarity which determines the interface shape. However, the fluctuations of $\mathrm{Ca}_{\perp}/\mathrm{Ca}$ differ among cases, as discussed below.

\begin{figure}[t]
	\centering
	\begin{subfigure}[b]{0.49\textwidth}
		\centering
		\includegraphics[width=\textwidth]{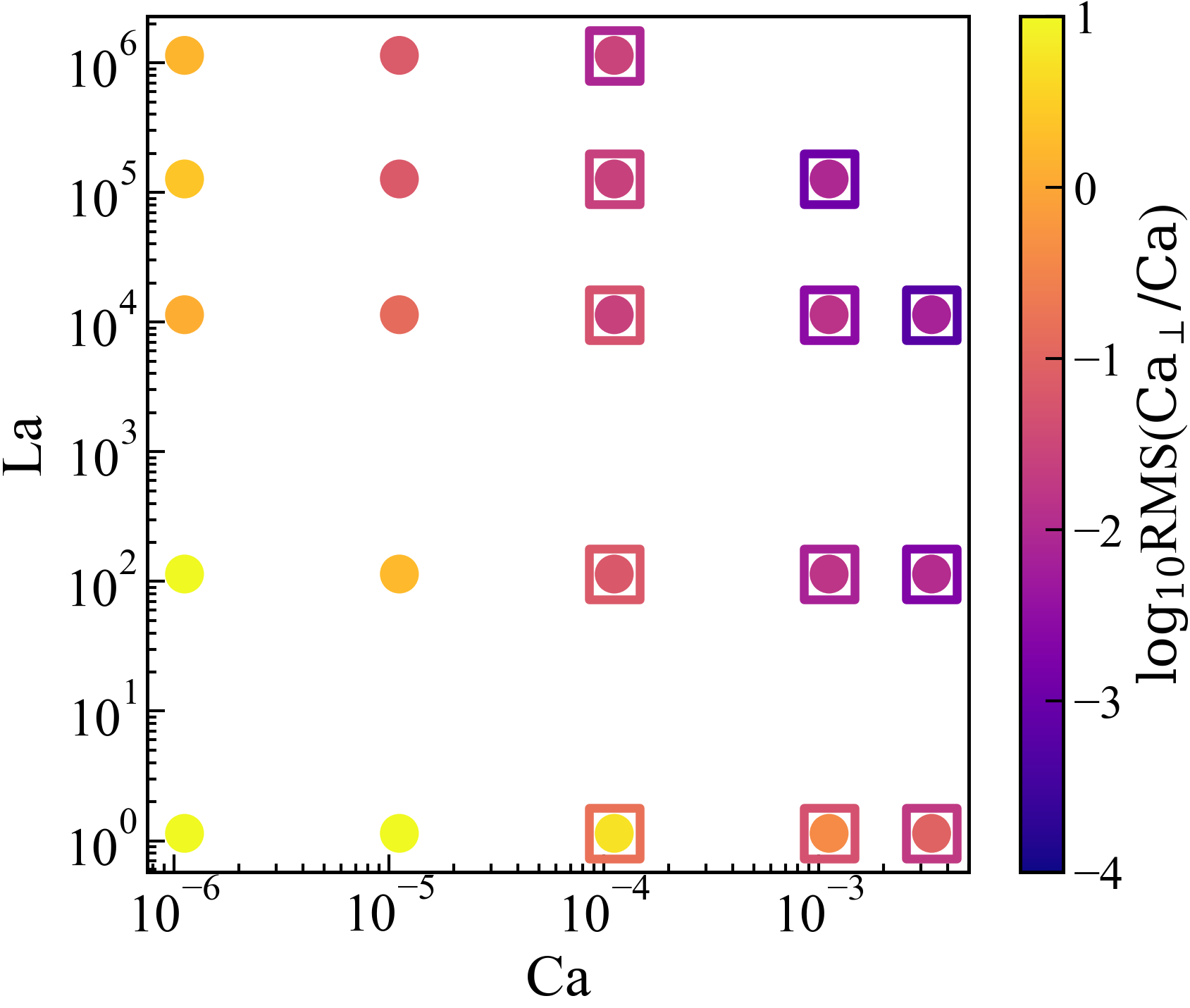}
		\caption{}
	\end{subfigure}
	\hfill
	\begin{subfigure}[b]{0.49\textwidth}
		\centering
		\includegraphics[width=\textwidth]{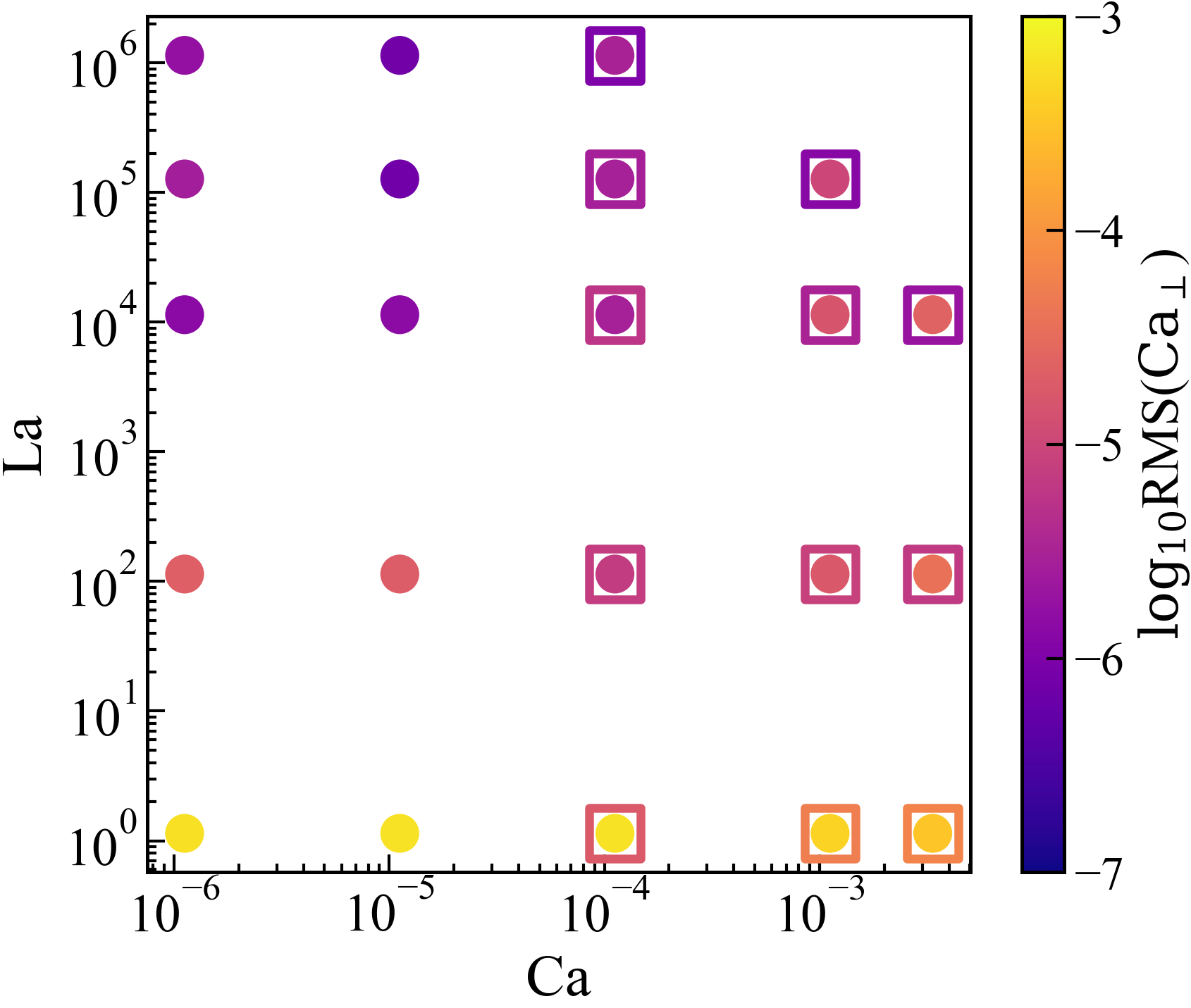}
		\caption{}
	\end{subfigure}
	\caption{RMS values of the fluctuations of (a) $\mathrm{Ca}_{\perp}/\mathrm{Ca}$ and (b) $\mathrm{Ca}_{\perp}$. The solid circles correspond to a Poiseuille-type inflow with moving contact lines (see Fig.~\ref{fig4-1}), whereas the hollow squares represent a uniform inflow without contact lines (see Fig.~\ref{figa-6}). The RMS values of $\mathrm{Ca}{\perp}/\mathrm{Ca}$ and $\mathrm{Ca}{\perp}$ are computed using Eq.~\eqref{a1}.}
	\label{figa-9}
\end{figure}

Fig.~\ref{figa-9} presents the RMS values of the $\mathrm{Ca}_{\perp}$ fluctuations (denoted by solid circles) over the parameter space shown in Fig.~\ref{figa-4}. When the Laplace number $\mathrm{La}$ is fixed, the fluctuation amplitude of  $\mathrm{Ca}_{\perp}/\mathrm{Ca}$ increases as $\mathrm{Ca}$ decreases (achieved by reducing $u_{\max}$), whereas the fluctuation amplitude of $\mathrm{Ca}_{\perp}$ itself remains nearly unchanged. This indicates that the wall-normal velocity fluctuations are largely independent of the inlet velocity, which is parallel to the solid wall. Such velocity fluctuations originate from the contact-angle enforcement, as discussed in Sec.~\ref{3.5.1}. In contrast, when the capillary number $\mathrm{Ca}$ is fixed, the fluctuation amplitudes of both $\mathrm{Ca}_{\perp}/\mathrm{Ca}$ and $\mathrm{Ca}_{\perp}$ decrease with increasing $\mathrm{La}$. Recall that increasing $\mathrm{La}$ corresponds to decreasing the fluid viscosity $\mu$, which in turn leads to an increase in the inlet velocity $u_{\max}$. Consequently, the ratio $\mathrm{Ca}_{\perp}/\mathrm{Ca}$ is reduced. However, the decrease in fluctuation amplitude eventually saturates at large $\mathrm{La}$, likely due to the  rapidly increasing Reynolds number.

%\begin{table}[t]
%\centering
%\caption{\color{blue}RMS of the fluctuations of the ratio $\mathrm{Ca}_{\perp}/\mathrm{Ca}$ for a Poiseuille-type inflow in an embedded microchannel with moving contact lines.}
%\begin{tabular}{c|ccccc}
%\hline
%\hline
%\multicolumn{1}{c|}{$\mathrm{Ca} \backslash \mathrm{La}$} 
%& $1.14$ 
%& $1.14\times10^{2}$ 
%& $1.14\times10^{4}$ 
%& $1.27\times10^{5}$ 
%& $1.14\times10^{6}$ \\
%\hline
%$1.12\times10^{-6}$ & 526    & 19.7   & 1.25    & 2.414   & 1.574  \\
%$1.12\times10^{-5}$ & 54.3   & 1.79   & 0.128   & 0.0668  & 0.0694 \\
%$1.12\times10^{-4}$ & 5.55   & 0.0644 & 0.0260  & 0.0259  & 0.0277 \\
%$1.12\times10^{-3}$ & 0.396  & 0.0157 & 0.0141  & 0.00918 & ---    \\
%$3.36\times10^{-3}$ & 0.0973 & 0.0112 & 0.00714 & ---     & ---    \\
%\hline
%\hline
%\end{tabular}
%\label{tab:Ca_Oh2}
%\end{table}

\begin{figure}[t]%% placement specifier
	\centering%% For centre alignment of image.
	\includegraphics[width=0.4\textwidth]{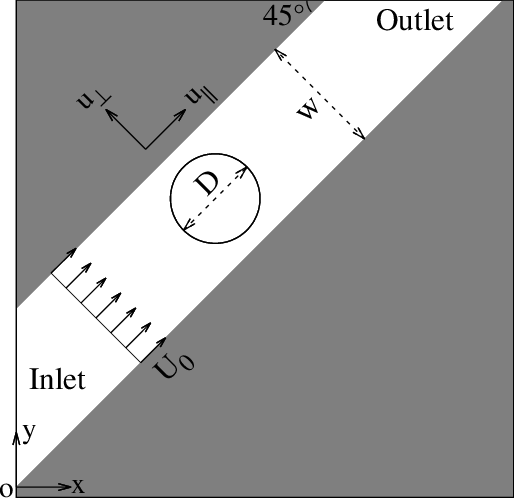}
	\caption{Schematic of a droplet advected by a uniform inflow in a straight microchannel.}\label{figa-6}
\end{figure}

To further isolate the influence of contact-angle enforcement on the fluctuations of $\mathrm{Ca}_{\perp}$, we consider a reference test case, as shown in Fig.~\ref{figa-6}, in which a droplet of diameter $D$ is advected by a uniform inflow with velocity $U_0$ through the same straight microchannel as in Fig.~\ref{fig4-1}. The channel width $w$ and the physical properties of both fluids are identical to those used in Sec.~\ref{3.5.1}. A free-slip boundary condition is imposed on the two embedded solid boundaries. In this configuration, the wall-normal velocity component $u_{\perp}$ should ideally vanish; thus, any nonzero $u_{\perp}$ represents spurious velocity arising solely from numerical errors. This test case serves as the embedded-boundary counterpart of the classical translating-droplet test used to evaluate spurious currents in the absence of solid walls~\cite{popinet2009accurate,ABADIE2015611,SAINI2025}.

\begin{figure}[t]
	\centering
	\begin{subfigure}[b]{0.49\textwidth}
		\centering
		\includegraphics[width=\textwidth]{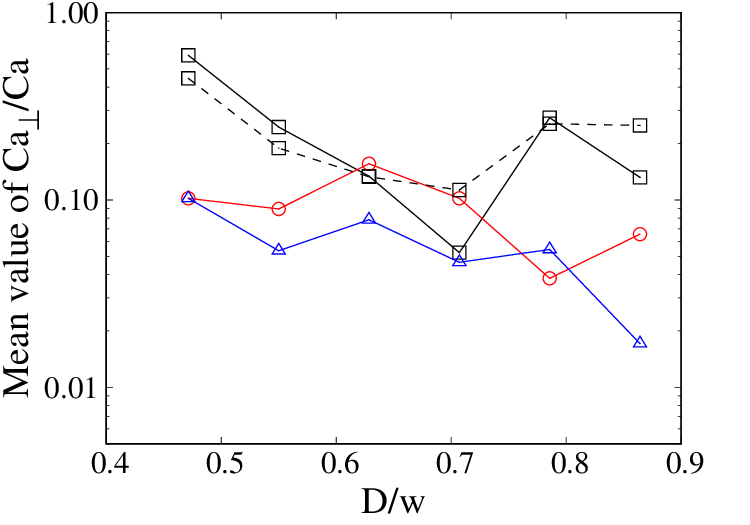}
		\caption{}
	\end{subfigure}
	\hfill
	\begin{subfigure}[b]{0.49\textwidth}
		\centering
		\includegraphics[width=\textwidth]{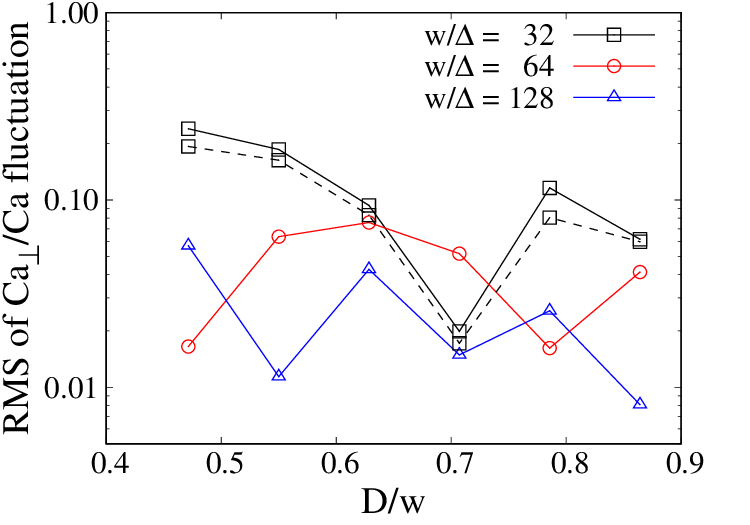}
		\caption{}
	\end{subfigure}
	\caption{Variations of $\mathrm{Ca}_{\perp}/\mathrm{Ca}$ with droplet size ($D/w = 0.47,\,0.55,\,0.63,\,0.71,\,0.79,$ and $0.86$) for different grid resolutions at $\mathrm{Ca}=1.12\times10^{-4}$ and $\mathrm{La}=1.14\times10^{4}$. The dashed line represents the case with a Poiseuille-type inflow and no-slip boundary conditions on the embedded solid walls at  $w/\Delta=32$.}
	\label{figa-7}
\end{figure}

Fig.~\ref{figa-7} presents the temporal mean value of $\mathrm{Ca}_{\perp}/\mathrm{Ca}$ and the RMS of its fluctuations (evaluated by Eq.~\eqref{a1}) at $\mathrm{Ca}=1.12\times10^{-4}$ and $\mathrm{La}=1.14\times10^{4}$ for different droplet sizes, where $\mathrm{Ca}=\mu U_0 /\sigma$. The variation of $\mathrm{Ca}_{\perp}/\mathrm{Ca}$ with droplet size reflects the wall effect on droplet translation. Overall, both the mean value and the fluctuation amplitude of $\mathrm{Ca}_{\perp}/\mathrm{Ca}$ decrease with grid refinement. For comparison, the droplet translation driven by a Poiseuille-type inflow with no-slip boundary conditions on the embedded solid walls (the 
setup shown in Fig.~\ref{fig4-1}) is also included, as indicated by the dashed line in Fig.~\ref{figa-7}. The fluctuations of $\mathrm{Ca}_{\perp}$ in the two configurations are found to be very similar.

\begin{figure}[tbp]
	\centering
	\begin{subfigure}[b]{0.49\textwidth}
		\centering
		\includegraphics[width=\textwidth]{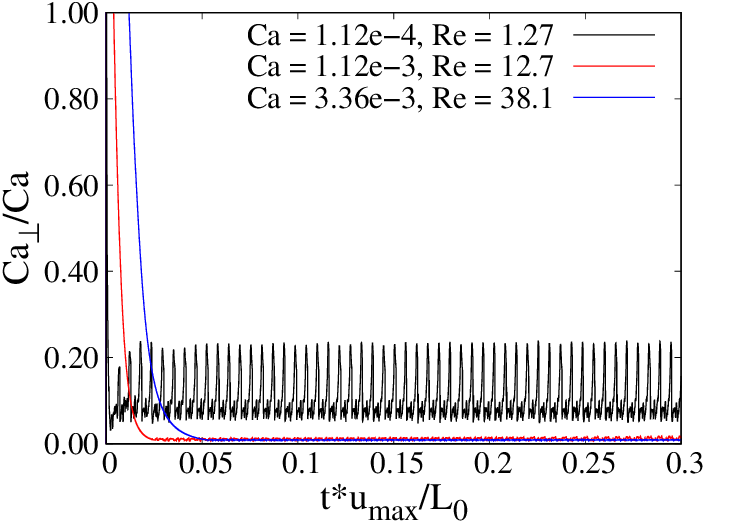}
		\caption{}
	\end{subfigure}
	\hfill
	\begin{subfigure}[b]{0.49\textwidth}
		\centering
		\includegraphics[width=\textwidth]{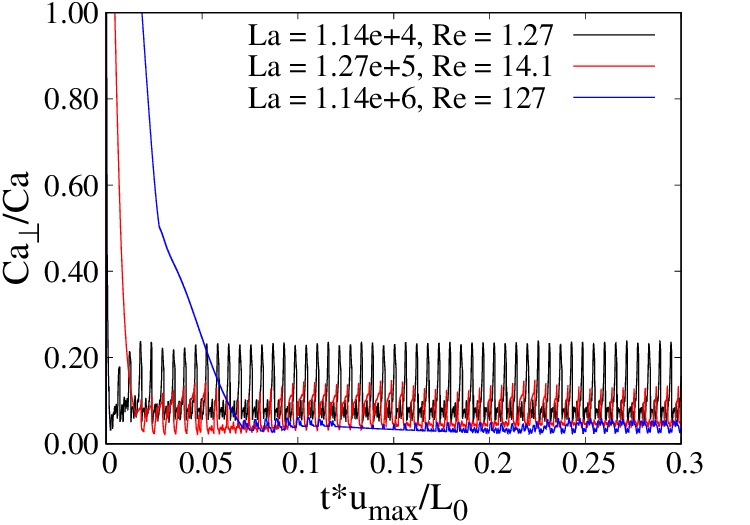}
		\caption{}
	\end{subfigure}
	\caption{Temporal evolution of the ratio $\mathrm{Ca}_{\perp}/\mathrm{Ca}$ for a uniform inflow in an embedded microchannel without contact lines: (a) different capillary numbers at $\mathrm{La} = 1.14\times10^{4}$ and (b) different Laplace numbers at $\mathrm{Ca}=1.12\times10^{-4}$.}
	\label{figa-8}
\end{figure}

In the following analysis, the droplet size is fixed at $D/w=0.707$ and the grid resolution at $w/\Delta=64$. We investigate the influence of the capillary number $\mathrm{Ca}$ and the Laplace number $\mathrm{La}$ on the spurious wall-normal velocity $u_{\perp}$. The results are shown in Fig.~\ref{figa-8}. In Fig.~\ref{figa-8}(a), both the magnitude and the fluctuation amplitude of $\mathrm{Ca}_{\perp}/\mathrm{Ca}$ at $\mathrm{Ca}=1.12\times10^{-4}$ are substantially larger than those in the other two cases, which is attributed to the significantly reduced inflow velocity $U_0$. In Fig.~\ref{figa-8}(b), both the magnitude and the fluctuation amplitude of $\mathrm{Ca}_{\perp}/\mathrm{Ca}$ decrease as $\mathrm{La}$ increases. Recall that increasing $\mathrm{La}$ corresponds to decreasing the fluid viscosity while increasing $U_0$ to maintain a fixed capillary number, thereby reducing the ratio $\mathrm{Ca}_{\perp}/\mathrm{Ca}$.

\begin{figure}[t]%% placement specifier
	\centering%% For centre alignment of image.
	\includegraphics[width=0.55\textwidth]{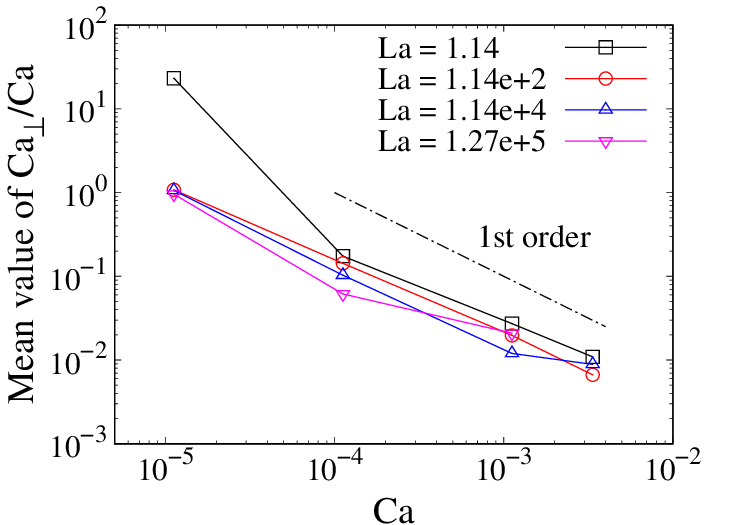}
	\caption{Temporal mean of the ratio $\mathrm{Ca}_{\perp}/\mathrm{Ca}$ as a function of the capillary number for a uniform inflow in an embedded microchannel without contact lines.}\label{figa-10}
\end{figure}

Fig.~\ref{figa-10} presents the variation of the temporal mean of $\mathrm{Ca}{\perp}/\mathrm{Ca}$ with the capillary number at different Laplace numbers. The mean value decreases approximately linearly as $\mathrm{Ca}$ increases, consistent with the trend observed in the translating-droplet test case without any solid walls \cite{popinet2009accurate}. In contrast, the mean value of $\mathrm{Ca}{\perp}/\mathrm{Ca}$ is largely insensitive to the Laplace number. Additionally, when $\mathrm{Ca}<1.12\times10^{-5}$, the magnitude of the spurious velocity characterized by $\mathrm{Ca}_{\perp}$ becomes comparable to $\mathrm{Ca}$, preventing the droplet from being transported by the imposed inflow.

Finally, the RMS values of the fluctuations of both $\mathrm{Ca}_{\perp}/\mathrm{Ca}$ and $\mathrm{Ca}_{\perp}$ are reported in Fig.~\ref{figa-9} as hollow squares. The variation trends with respect to $\mathrm{Ca}$ and $\mathrm{La}$ are consistent with those obtained in the presence of moving contact lines (denoted by solid circles). For most cases with $\mathrm{Ca}>1.12\times10^{-4}$, the fluctuation amplitudes of $\mathrm{Ca}_{\perp}/\mathrm{Ca}$ in the configuration without contact lines are smaller than those with moving contact lines. However, when $\mathrm{Ca}<1.12\times10^{-5}$, the spurious velocity becomes comparable in magnitude to the imposed inflow, preventing the droplet from being transported. Consequently, the corresponding RMS values are not reported. This behavior is consistent with that previously observed for a translating droplet in the absence of solid walls, as shown in Fig.~\ref{fig4-3}(b).

\bibliographystyle{elsarticle-num}
\bibliography{refs}

\end{document}